  \documentclass[12pt,preprint]{aastex}

\slugcomment{}

\shorttitle{Globular Cluster Systems}
\shortauthors{Mar\' \i n-Franch, A.}

\begin{document}

\title{Globular cluster systems in elliptical galaxies of Coma}

\author{A. Mar\' \i n-Franch}
\affil{Instituto de Astrof\' \i sica de Canarias, E-38205 La Laguna,
Tenerife, Spain}
\email{amarin@ll.iac.es}

\and

\author{A. Aparicio}
\affil{Instituto de Astrof\' \i sica de Canarias, E-38205 La Laguna,
Tenerife, Spain}
\email{aaj@ll.iac.es}

\begin{abstract}

Globular cluster systems of 17 elliptical
galaxies have been studied in the Coma cluster of galaxies. Surface-brightness fluctuations have been used to determine 
total populations of globular clusters and specific frequency ($S_N$)
 has been evaluated for each individual galaxy. Enormous differences in $S_N$ between similar galaxies are found. In particular, $S_N$ results vary by an order of magnitude from galaxy to galaxy. Extreme cases are the following: 
a) at the lower end of the range,
 NGC 4673 has $S_N = 1.0 \pm 0.4$, a surprising value for an elliptical 
galaxy, but typical for spiral and irregular galaxies; b) at the upper 
extreme, 
MCG +5 $-$31 $-$063 has $S_N = 13.0 \pm 4.2$ and IC 4051 $S_N = 12.7 \pm 3.2$, and are more likely to 
belong 
to supergiant cD galaxies than to ``normal" elliptical galaxies. 
Furthermore,  NGC 4874, the central supergiant cD galaxy of the Coma 
cluster, also exhibits a relatively high specific frequency 
($S_N = 9.0 \pm 2.2$). The other galaxies studied  have $S_N$ 
in the range [2, 7],  the mean value being $S_N = 5.1$. No single 
scenario seems to account for the observed specific frequencies, 
so the history of each galaxy must be deduced individually by 
suitably combining  the different models ({\it in situ}, mergers, 
and accretions). The possibility that Coma is formed by several subgroups is also considered. If only the galaxies of the main subgroup defined by \citet{GM01} are used, a trend in $S_N$ arises in the sense of $S_N$ being bigger in higher density regions. This result needs further confirmation.

\end{abstract}


\section{Introduction}

Globular clusters (GCs) are thought to be among the oldest objects 
in the
Universe; so, although they represent only a small fraction of the 
total
luminosity of the galaxy hosting them, they provide useful information 
about
the galaxy formation process and its initial evolutionary stages. By 
studying
globular cluster systems (GCS), the memory of the host galaxy is probed.

A fundamental, widely used parameter describing a GCS is the specific
frequency, $S_N$, introduced by \citet{HB81}. $S_N$ is the number of GCs
normalized to an absolute magnitude for the parent galaxy of
$M_V^{\rm TOT}=-15$ mag:

\begin{equation}
S_N=N_{\rm GC}^{\rm tot} 10^{0.4(M_V^{\rm TOT} + 15)},
\end{equation}
where $N_{\rm tot}$ is the total number of GCs in the galaxy. $S_N$ gives the
number of GCs per unit luminosity of the host galaxy. Spiral and irregular
galaxies have typical values of $S_N \le 1$ \citep{BH82}. 
Dwarf \citep{D96,M98},
elliptical \citep{HB81}, and lenticular galaxies \citep{KW98,C99} have $S_N$
between 2 and 6, where larger values seem to correspond to galaxies located in
high-density environments \citep{H91}. Giant cDs located near the centers of
rich clusters of galaxies have the largest specific frequency, $S_N \sim 10 -
20$ (Harris \& van den Berg 1981; Harris, Pritchet, \& McClure 1995; Blakeslee
\& Tonry 1995; Bridges et al. 1996), with some exceptions \citep{K96}.

The origin of the observed values of $S_N$ is not clear. A basic question is
whether $S_N$ depends on the properties of the host galaxy and/or the
environment. If only elliptical galaxies are considered, environmental
density seems to play a fundamental role in $S_N$ \citep{H91}. These results, 
which are based on the compilation of several studies of elliptical galaxies 
closer than the Virgo cluster,
suggest that the formation of GCs has been two or three times more efficient 
in
rich environments than in poor ones. On the other hand, the luminosity of the
host galaxy also has an influence on the properties of the GCS 
(Harris, Harris, \& McLaughlin 1998).

\citet{F82} and \citet{M87} have suggested that in clusters of galaxies a 
population of
intergalactic GCs (IGCs) should exist. A galaxy could enlarge its GC population via gravitational capture of IGCs. If they are distributed according to the gravitational
potential of the whole galaxy cluster, then the influence of the environment on $S_N$ is justified \citep{W87,W95}. 
On the other hand, \citet{B97} suggested that $S_N$
could be strongly influenced by gas removal by tidal disruption during the formation of the cluster. This idea was developed further by Blakeslee, Tonry, \& Metzger (1997). The star formation would be inhibited as a result of the gas loss, increasing $S_N$. Moreover, \citet{H98} argued that explaining
the observed $S_N$ in M87 via gravitational capture of IGCs, requires assuming an unrealistic spatial distribution for intergalactic GCs. 

The origin of the GCSs themselves in giant elliptical galaxies is still not understood. One of the main problems in understanding this phenomenon is the lack of data. The difficulty in observing distant GCSs is due to the low resolution of available telescopes and to the low brightness of the individual GCs, which are too faint to be detected even at moderate distances. At $\sim 100$ Mpc (the distance of the Coma galaxy cluster), only the brightest GCs are visible from ground-based observations (\citet{H87} and \citet{TV87} performed the first studies of GCSs in Coma); high-resolution telescopes (such as the {\it Hubble Space Telescope} ({\it HST\/}), see for example \citet{H00}, \citet{K00}) and long exposure times are required to detect them directly. 
But the surface-brightness fluctuations (SBF) technique, which we use in
this paper, allows us to detect GCSs in elliptical galaxies 
at
distances of the order of 100 Mpc, using small to intermediate-size,
ground-based telescopes and reasonable exposure times \citep{BT95}.

The SBF technique was introduced by \citet{TS88} with the aim of measuring
distances. Comparing the fluctuation signals produced by the stellar
population of a galaxy with SBF measurements of nearby galaxies which have externally calibrated distances, accurate estimates of 
distances
can be obtained up to $\sim$ 40 Mpc \citep{T00,T01}. For smaller
distances, undetected GCs contribute as a perturbation of the total 
fluctuation
signal of the galaxy. But in the case of more distant galaxies the
fluctuation signal produced by the GCs, which are mostly unresolved, 
dominates and the contribution of stars becomes negligible 
(Wing et al. 1995; Blakeslee \& Tonry 1995; Blakeslee, Tonry, 
\& Metzger 1997;
Blakeslee 1999).

In this paper, we use the  SBF technique
to evaluate the GCS properties of elliptical
galaxies in the Coma cluster. Coma is a good laboratory for studying GCSs,
because it is a rich cluster and its ellipticals have a wide range of
luminosities and environmental densities. 
After describing the SBF technique and its use in obtaining 
GCS properties (\S \ref{SBF}), we present a test for evaluating the
capacity of the technique (\S \ref{test}), which we then apply to the
Coma cluster galaxies (\S \ref{result}).

\section{Observations and Data Reduction} \label{data}

This study is based on observations done in 2000 April 25 and 27 with the 
2.5 m
INT at the  Roque de los Muchachos Observatory on La Palma, using
the Wide Field Camera, which consists of four 2048 $\times$ 4100 pixel
chips with a scale of 0.333 $\arcsec$/pix, providing a field of view per
chip of 11.38$\arcmin$ $\times$ 22.75$\arcmin$. The total field of view is 
$\sim34\arcmin$ on a side. 

\begin{figure}
\epsscale{0.5}
\plotone{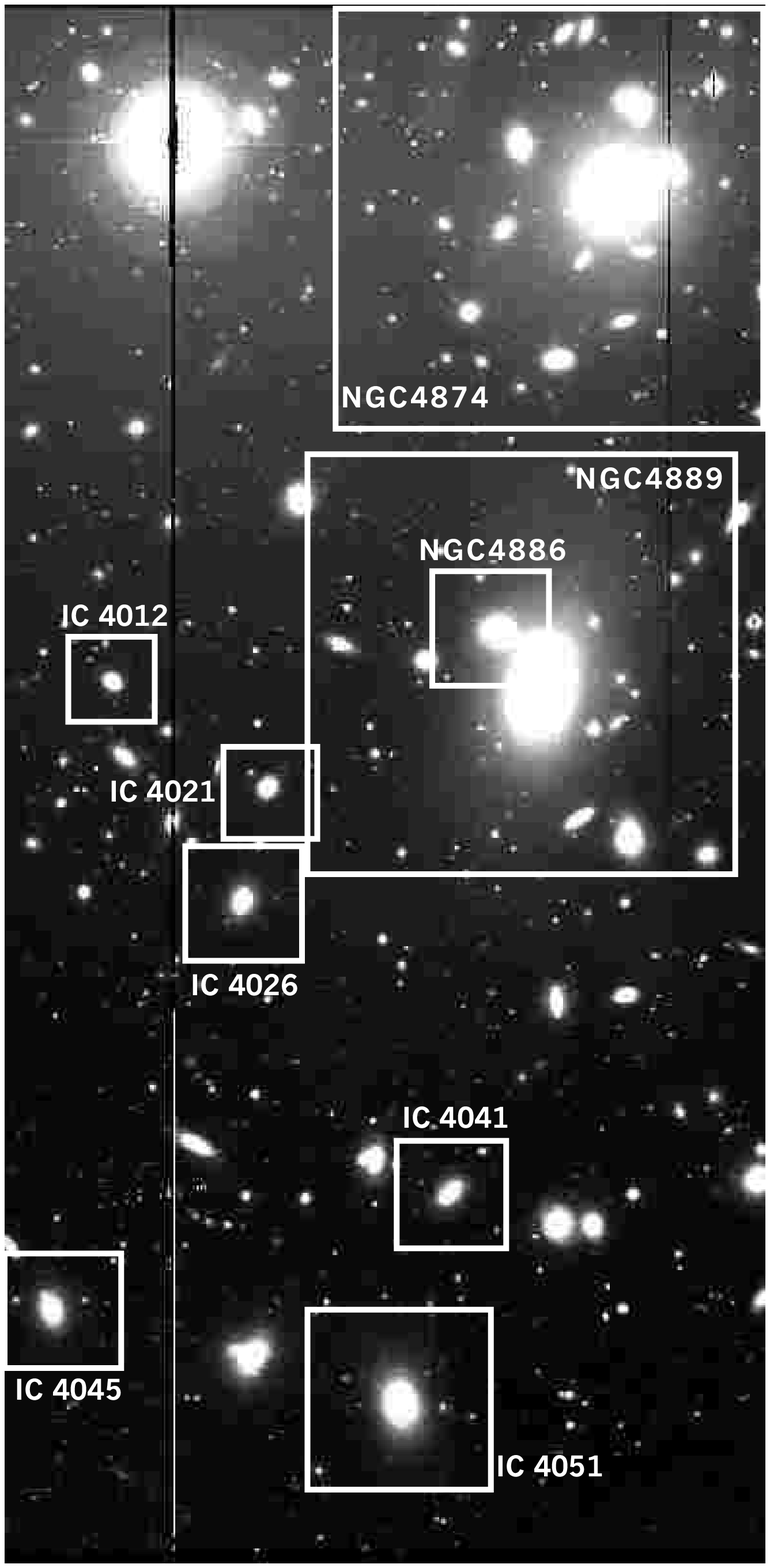} 
\caption{Central chip of the CCD mosaic in the target 
field RC-1. The nine galaxies studied in this chip are 
labeled. \label{chip4}}
\epsscale{1}
\end{figure}

In Table \ref{coords}, the location of the target fields and all the  
galaxies 
studied in each field are listed. Column 1 gives the name adopted 
in this article
 for each target field, columns 2 and 3 give the coordinates of the field and 
column 4 gives the galaxies analyzed in each field. Twelve exposures of 300 s 
with the Sloan $R$ Filter were taken for each field, except for the central 
field of Coma (RC-1) where thirteen exposures of 300 s were taken; this 
resulted
 in a total integration time of 3600 s (or 3900 s for RC-1).  In Table~\ref{obs}, a summary of the observations is given. Column 1 lists the field names, column 2 the date of the observation, columns 3 and 4 
give the exposure times and the used filter, and column 5 gives the 
seeing, which is between 1$\arcsec$ and 1.2$\arcsec$ in all cases. The bias level 
was subtracted from the individual frames using the overscan region of 
the CCD. The data were then flattened with twilight sky flats. The final 
image was produced by calibrating the individual frames, dithering 
between exposures and calculating an average of the frames. In 
Fig.  \ref{chip4}, the central chip of the CCD is shown for the 
target field RC-1, where nine studied galaxies are marked. This field corresponds to the center of the Coma cluster. 

Photometric calibration was performed for the second night. The observations of the first night were then calibrated comparing the field RC-1 observed in both nights.
The extinction and photometric calibration were calculated using five measurements of 23 \citet{L92} standard stars. The equation transforming instrumental to standard magnitudes is

\begin{equation} \label{photocalibr}
R-r+AX=a_i+b_i\times(V-R),
\end{equation}
where $R$ is the magnitude in the standard system, $r$ is the instrumental magnitude, $X$ is the airmass (all the observations of target objects have airmass smaller than 1.1) and $A$ is the extinction, for which a value $A=0.145$ was obtained. $(V-R)$ is the colour index of the program objects. Since no observations were done in the $V$ Filter, a typical value for the colour index of GCs of $(V-R)=0.46$ was adopted \citep{A94}. $a_i$ and $b_i$ are the zero point and the color term for each chip (i=1, 2, 3 and 4). Our results for $a_i$ and $b_i$ are listed in Table \ref{calibr1}. The sigma of the photometric calibration varies between 0.01 and 0.02 mag. Calibration results for each individual galaxy are shown in Table \ref{calibr2}, where $m^{*}_{1}$ is the magnitude of an object yielding one unit of flux per unit time. This magnitude will be used later. The sigma of $m^{*}_{1}$ represents the sigma of the photometric calibration. Finally, as Coma is nearly perpendicular to the Galactic plane, interstellar extinction is neglected.

\section{Surface Brightness Fluctuations: the Technique} \label{SBF}

The concept of surface-brightness fluctuations was  introduced by 
\citet{TS88},
who noted that in the surface
photometry of a galaxy far enough to remain unresolved, a 
pixel-to-pixel
fluctuation is observed due to the Poisson statistics of the spatial
distribution of stars, GCs, background galaxies, foreground stars, etc. This
is correct for a galaxy with no absorption or emission due to dust or gas; so
this technique would only be applicable to early-type galaxies. The variance
of the fluctuation depends on the stellar population, the GCS, background
galaxies, foreground star, and, of course, the distance.

\subsection{Sources of Variance:  GCS Properties} \label{sources} 

The following treatment has been described by \citet{BT95}. It can be assumed
that the total pixel-to-pixel variance of an image ($\sigma^{2}_{\rm TOT}$) is
the sum of all the independent contributions due to the different components
in the image, such as the stellar population ($\sigma^{2}_{\rm SP}$), GCs
($\sigma^{2}_{\rm GC}$), background galaxies ($\sigma^{2}_{\rm BG}$), 
foreground
stars ($\sigma^{2}_{\rm FS}$), read-out noise ($\sigma^{2}_{\rm ro}$) 
and photon
shot noise ($\sigma^{2}_{\rm ph}$):

\begin{equation} \label{sigma-tot}
\sigma^{2}_{\rm TOT}=\sigma^{2}_{\rm SP}+\sigma^{2}_{\rm 
GC}+\sigma^{2}_{\rm BG}+\sigma^{2}_{\rm FS}+\sigma^{2}_{\rm ro}+
\sigma^{2}_{\rm ph}
\end{equation}

The pixel-to-pixel variance produced by a class of object can be evaluated as
the second moment of the luminosity function of that object population. In
general, the brightest individuals could be resolved in the image. For the
analysis of the SBF, these objects must be masked. If all sources brighter
than a limiting flux ($f_{\rm lim}$) are masked in the image, then the 
variance
produced by the remaining part of the population is:

\begin{equation} \label{sigma}
\sigma^{2}_{\rm POP}=\int_0^{f_{\rm lim}} n(f) f^2 df,
\end{equation}
where $n(f)$ is the luminosity function; i.e., the number of population
objects per unit flux per pixel. We can put this equation in terms of
magnitudes via the relationship

\begin{equation}
m=-2.5 \log(f) + m^*_1,
\end{equation} 
$f$ being the flux ($e^-$ s$^{-1}$ pix$^{-1}$) and $m^{*}_{1}$ the magnitude of an object yielding one unit of flux per unit time.

If $n(f)$ is known, the variance produced by the non-masked part of the
population can be evaluated. In the case of GCs, a Gaussian shape
for the luminosity function (GCLF) can be assumed \citep{H91}:

\begin{equation} \label{gclf}
n_{\rm GC}(m)=\frac{N_{\rm GC}}{\sqrt{2 \pi} \sigma} e^{{-(m-m^0)^2}/{2 
\sigma^2}},
\end{equation}
where $N_{\rm GC}$ is the total number of GCs per pixel. Substituting
eq. \ref{gclf} into eq. \ref{sigma}, the variance due to the non-masked
globular clusters can be expressed as:

\begin{equation} \label{sigma-gc}
\sigma^2_{\rm GC}=\frac{N_{\rm GC}}{\sqrt{2 \pi} \sigma} 10^{0.8 m^*_1} 
\int_{m_{\rm c}}^{\infty} e^{{-(m-m^0)^2}/{2 \sigma^2}} e^{-0.8 \ln(10) m} 
dm,
\end{equation}
where $m_{\rm c}$ is the magnitude corresponding to the limiting flux $f_{\rm lim}$.

It can be seen that the variance produced by the GCs is proportional 
to their
total surface density. In this way, assuming values for $m^{0}$ and 
$\sigma$
in the GCLF, and determining the variance produced by GCs, $\sigma^2_{\rm 
GC}$
from the image analysis, their surface density can be obtained. In galaxies 
with large angular size, the spatial
structure of the GCS can be also evaluated by dividing the galaxy into 
rings and
measuring the GCs' variance in each one.

What is directly obtained from the analysis of the image is
$\sigma^{2}_{\rm TOT}$. Hence a treatment of the remaining sources of variance 
is
required. First, at the distance of Coma, using eq. \ref{sigma-gc}
 and with
the help of the Padua stellar evolution library (see \citet{Be94}),
it can be shown that the ratio $\sigma^{2}_{\rm SP}$/$\sigma^{2}_{\rm
 GC}$ is nearly
zero, so $\sigma^{2}_{\rm SP}$ can be neglected in eq.
 \ref{sigma-tot}. Also,
as Coma is nearly perpendicular to the Galactic plane, foreground
stars are very few, and $\sigma^{2}_{\rm FS}$ can be neglected. Hence,

\begin{equation}
\sigma^{2}_{\rm TOT} \approx \sigma^{2}_{\rm GC}+\sigma^{2}_{\rm 
BG}+\sigma^{2}_{\rm ph}+\sigma^{2}_{\rm ro}.
\end{equation}

As we will see later, the SBF technique involves  spectral analysis of 
the
signal. This provides two quantities: $P{_1}$ and $P{_0}$. $P{_0}$ is the
total PSF-convolved variance, produced by all objects whose spatial flux
distribution is convolved with the PSF. $P{_1}$ is the non-PSF-convolved
variance. As a result,

\begin{equation} \label{P0}
P_0\approx \sigma^{2}_{\rm GC}+\sigma^{2}_{\rm BG} 
\end{equation}
and
\begin{equation}
P_1=\sigma^{2}_{\rm ro}+\sigma^{2}_{\rm ph}.
\end{equation}
Consequently, the only contribution to $\sigma^{2}_{\rm TOT}$ we have to 
be concerned
with is $\sigma^{2}_{\rm BG}$. Once this is removed, $\sigma^{2}_{\rm GC}$ 
will
be obtained from $P_0$. 

The following law can be assumed for the luminosity function of 
background
galaxies (BGLF): 

\begin{equation} \label{bglf}
n_{\rm BG}(m)=A \times N_{\rm BG} 10^{\gamma m},
\end{equation}
where $A$ is a constant for normalization, $N_{\rm BG}$ is the total number
 of galaxies per pixel and $\gamma$ is the slope of the magnitude 
distribution. \citet{T88} found $\gamma \approx 0.39$ in the $R$ 
filter. Using eq. \ref{sigma}, we find

\begin{equation} \label{sigma_bg}
\sigma_{\rm BG}^2=\frac{A \times N_{\rm BG}}{\ln10(0.8-\gamma)}10^{\gamma
 m_{\rm c}}10^{0.8(m_1^*-m_{\rm c})}.
\end{equation}

Once the parameters $A$, $N_{\rm BG}$ and $m_{\rm c}$ are estimated from an 
analysis of the image, as we will see later, and $m_1^*$ from the photometric 
calibration, $\sigma_{\rm BG}^2$ can be calculated.

\subsection{Measuring \boldmath $\sigma^{2}_{\rm GC}$\unboldmath}

As we have seen, because stars, background galaxies, and GCs are convolved 
with the PSF, while the read-out and photon shot noise are not, if a  
power spectrum of the image is computed,  the non-convolved variance can 
be separated from the remaining terms. The power spectrum of the image, 
$P(k)$, has the form

\begin{equation} \label{power}
P(k)=P_1+P_0 E(k),
\end{equation}
where $E(k)$ is the power spectrum of the PSF convolved with the window 
function (which is unity in all the regions we want to study and zero 
in the rest of the image), $P_0$ is the PSF-convolved variance 
($P_0$ = $\sigma^{2}_{\rm SP}$+$\sigma^{2}_{\rm GC}$+$\sigma^{2}_{\rm 
BG}$+$\sigma^{2}_{\rm FS}$+...), and $P_1$ is the non-PSF-convolved 
variance ($P_1$ = $\sigma^{2}_{\rm ro}$+$\sigma^{2}_{\rm ph}$).

Before performing the power spectrum of the image, a preliminary image 
treatment has to be done.

\subsubsection{Preliminary image treatment}

After the initial image reduction was done, and because there are 
some galaxies located in each chip of the CCD mosaic, different 
regions of the chips were selected, each  containing a galaxy (see 
Fig. \ref{chip4}) that was treated individually. In all cases, a 
smooth fit of the galaxy profile was performed using the IRAF 
package ISOPHOTE. An iterative method was used: first, bright 
objects that could affect the galaxy profile fit were masked. 
The task ELLIPSE was used to fit this profile, and
 the result was subtracted from the original image in order to 
obtain a map of the bright objects alone. These were then subtracted 
from the initial image and a second fit of the galaxy profile was made. 
This procedure was repeated until the precision of the final fit was 
good enough. 

Once the final galaxy profile was performed
 and subtracted, a residual zero mean image is obtained where the 
PSF-convolved variance can be measured. In this image the power spectrum 
is only perturbed at low wave numbers (Jensen, Tonry, \& Luppino 1999), 
so this part of the power spectrum will not be taken into account in
 our analysis. 

Using the DAOPHOT task DAOFIND, all the objects 4$\sigma$ above the zero
 mean of the residual image are selected. Those  brighter than some 
limiting magnitude ($m_{\rm c}$) were masked out creating a window 
function whose value
 is zero where the bright object is located and 
unity
 in the rest of the image. The magnitude $m_{\rm c}$ must be 
chosen appropriately. Note that if a very bright $m_{\rm c}$ is 
used, $P_0$ is dominated by the non-masked brightest objects (since 
it is the second moment of the luminosity function of the population 
which is at work). In our study, brightest objects are background
 galaxies, so the GC contribution is masked out. Because the CG
 contribution will be evaluated, as many bright objects as possible
 must be masked out. On the other hand, if $m_{\rm c}$ is chosen too 
faint, the results could be affected by incompleteness,
 resulting in an incorrect estimate of the unresolved background
 galaxy contribution to $P_0$. An analysis of this question has to 
be made in each case. In this article, we choose $m_{\rm c} = 
23.5$. 

In large-angular-size galaxies, the spatial distribution of the GCS 
can be analyzed. In those cases, the set of ring-shaped regions in 
which the $P_0$ will be measured can be defined making use of the 
galaxy profile fit. We can select different isophotes to define the 
annular sections, in this way, elliptical annular regions are obtained 
at different distances from the center of the galaxy. Multiplying the 
previous window function with each defined annular region, a set of new 
window functions is obtained, one for each ring that we shall be studying. 

\subsubsection{Measuring $\sigma^{2}_{\rm GC}$}

Multiplying the window function of a selected region by the residual 
image, the residual masked image of the region is obtained. After this, 
the power spectrum of the residual masked image is computed. The result 
is a two-dimensional power spectrum, which is radially averaged in order 
to obtain the one-dimensional power spectrum of the selected region. 
This has the form expressed in equation \ref{power}. In order to evaluate
 $P_0$, the power spectrum of the PSF convolved with the window function 
of the region is needed. But $E(k)$ can be approximated by the power spectrum
 of the PSF alone, $P_{\rm PSF}(k)$, with a negligible error \citep{JTL98}, 
so eq. \ref{power} transforms into

\begin{equation} \label{power2}
P(k)=P_1+P_0 P_{\rm PSF}(k).
\end{equation}

Using an empirical form for the PSF, obtained from some bright, isolated 
stars in the same image being analyzed, and fitting the power spectrum of 
the images with eq. \ref{power2}, the quantities $P_0$ and $P_1$ can 
be evaluated. In 
large-angular-size galaxies, the spatial structure of the GCS can be 
deduced from the radial dependence of $P_0$. In the case of smaller
 galaxies, where we cannot measure radial dependences, only the total
 number of GCs is evaluated.

\section{Testing the Technique Using Synthetic Images} \label{test}

\begin{figure}
\plotone{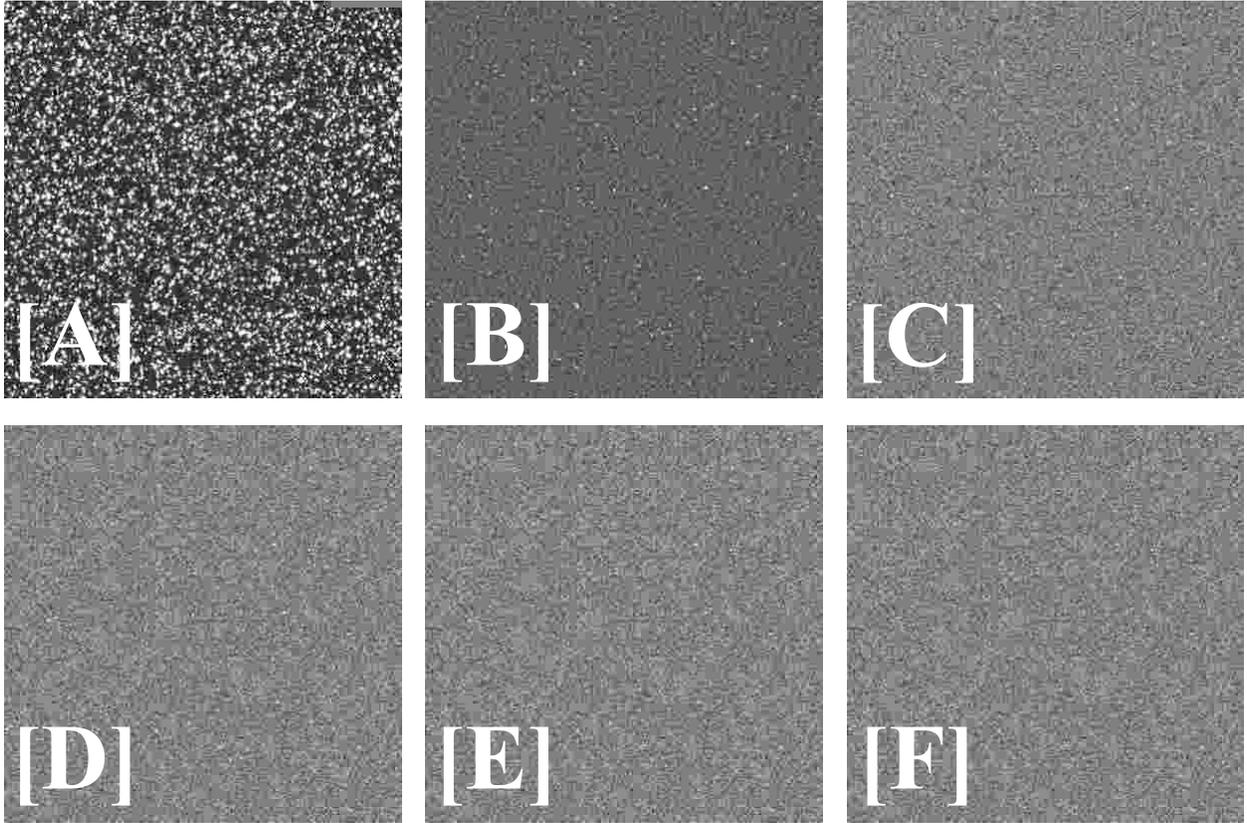}
\caption{Synthetic images generated for the internal 
consistency test  of the SBF to measure GCS properties. Image (a) 
is noise-free. Noise increases from left to right and from top 
to bottom. See text and Table \ref{testing} 
for details. \label{test_imag}}
\end{figure}

Putting into practice what we have presented in the preceding sections, 
in this section we will make a consistency test to check our procedure
for measuring the power spectrum normalization. The influence of 
the noise level in the images is also tested. Synthetic images were 
constructed with this purpose in the following manner. The magnitudes
 of the GC were randomly calculated according to a Gaussian distribution. 
The parameters of the GCLF were $\sigma=1.4$ and $m^0=20$. This is 
equivalent to supposing that the GCS is at a distance of 4 Mpc. At 
this distance, GCs have a stellar appearance, so simulating them with 
an empirical PSF is adequate. Using an empirical PSF and the DAOPHOT 
task ADDSTAR, we generated 10\,000 artificial GCs uniformly distributed 
across a 1000$^2$ pixels image. Six different values for the 
non-convolved variance were used (Fig. \ref{test_imag}) and the 
SBF technique was applied to the final synthetic images in order to 
recover the total number of GCs. Since all the objects in the image are 
GCs, it is not necessary to mask the brightest sources, so we used 
$m_{\rm c}=-\infty$ and  $m_1^*=25$ in
equation
 \ref{sigma-gc}. In this way we obtain $\sigma_{\rm GC}^2=0.278 
N_{\rm TOT}$, where N$_{\rm TOT}$ is the estimate for total number 
of GCs in the synthetic image. Finally, the SBF technique will produce 
for each image the values of $P_1$=$\sigma_{\rm noise}^2$ and 
$P_0$=$\sigma_{\rm GC}^2$.

\begin{figure}
\plotone{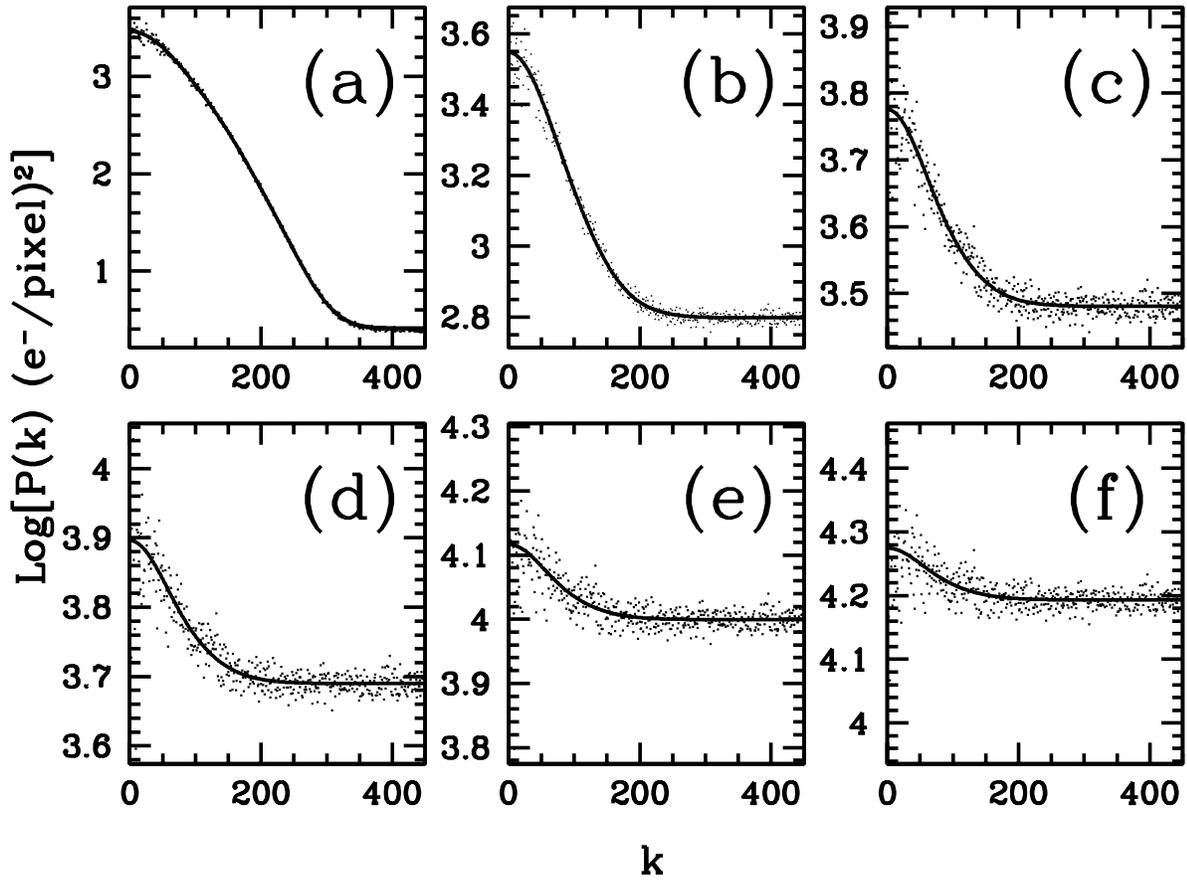}
\caption{SBF analysis of the synthetic images with 
different noise levels. Points represent the power spectrum of
 each image, and the solid line is the SBF fitting 
(eq. \ref{power2}). See Table~\ref{testing} for 
details. \label{testing_fig}}
\end{figure}

In Fig. \ref{testing_fig}, the power spectrum of the synthetic 
images for the six different noise levels are presented. 
Table \ref{testing} lists the properties of the synthetic images 
(columns 1 to 3) and the results of the SBF analysis (columns 4 to
 6). The test goes from images with no noise at all 
(image {\it a\/}) to completely noise-dominated images (images 
{\it e} and {\it f\/}).
 Here, $\sigma_{\rm noise}^2$ is the input noise level, 
$N_{\rm GC}^{\rm in}$ is the number of GCs introduced in the 
image, $P_0$ and $P_1$ are the SBF results for each image, and 
$N_{\rm GC}^{\rm out}$ is the total population of GCs estimated 
using equation \ref{sigma-gc}.

It is very interesting, even in the images with extremely high 
noise variance, that SBF  produces good results. In all cases, 
there is excellent agreement between the parameters of the input 
images and those recovered by the SBF technique, both for $\sigma_{\rm 
noise}^2$ and $\sigma_{\rm GC}^2$. We only observe a small systematic
 excess in the estimated number of GCs. This excess could be due to 
small differences between the real input GCLF (which includes random 
effects) and the exact Gaussian-shaped GCLF used to evaluate the final 
result in equation \ref{sigma-gc}. In order to check this, we 
calculated $\sigma_{\rm GC}^2$ directly from the magnitudes of all 
the 10\,000 GCs added to the images in the following way:

\begin{equation}
\sigma_{\rm GC}^2= \sum_{i=1}^{10\,000} f_i^2 = 
\sum_{i=1}^{10\,000} 10^{-0.8(m_i-25)}.
\end{equation}

The result was $\sigma_{\rm GC}^2$ = 2926 ($e^-$/pixel)$^2$, 
in full agreement with the obtained SBF results (column 5 in 
Table \ref{testing}). With the former test we have shown that 
SBF is a very powerful, selfconsistent technique, 
and that very good results can be obtained even with completely 
noise-dominated images, where traditional techniques do not 
detect any object.

\section{Results} \label{result}

\subsection{Measuring the BGLF and \boldmath $\sigma^2_{\rm 
BG}$\unboldmath} \label{sigmaBG}

The following treatment is the same for all studied galaxies,
 but with the aim of illustrating the methodology, we first present 
the procedure for the galaxy NGC 4874 in detail and then the results 
for the rest of galaxies.

\begin{figure} 
\plotone{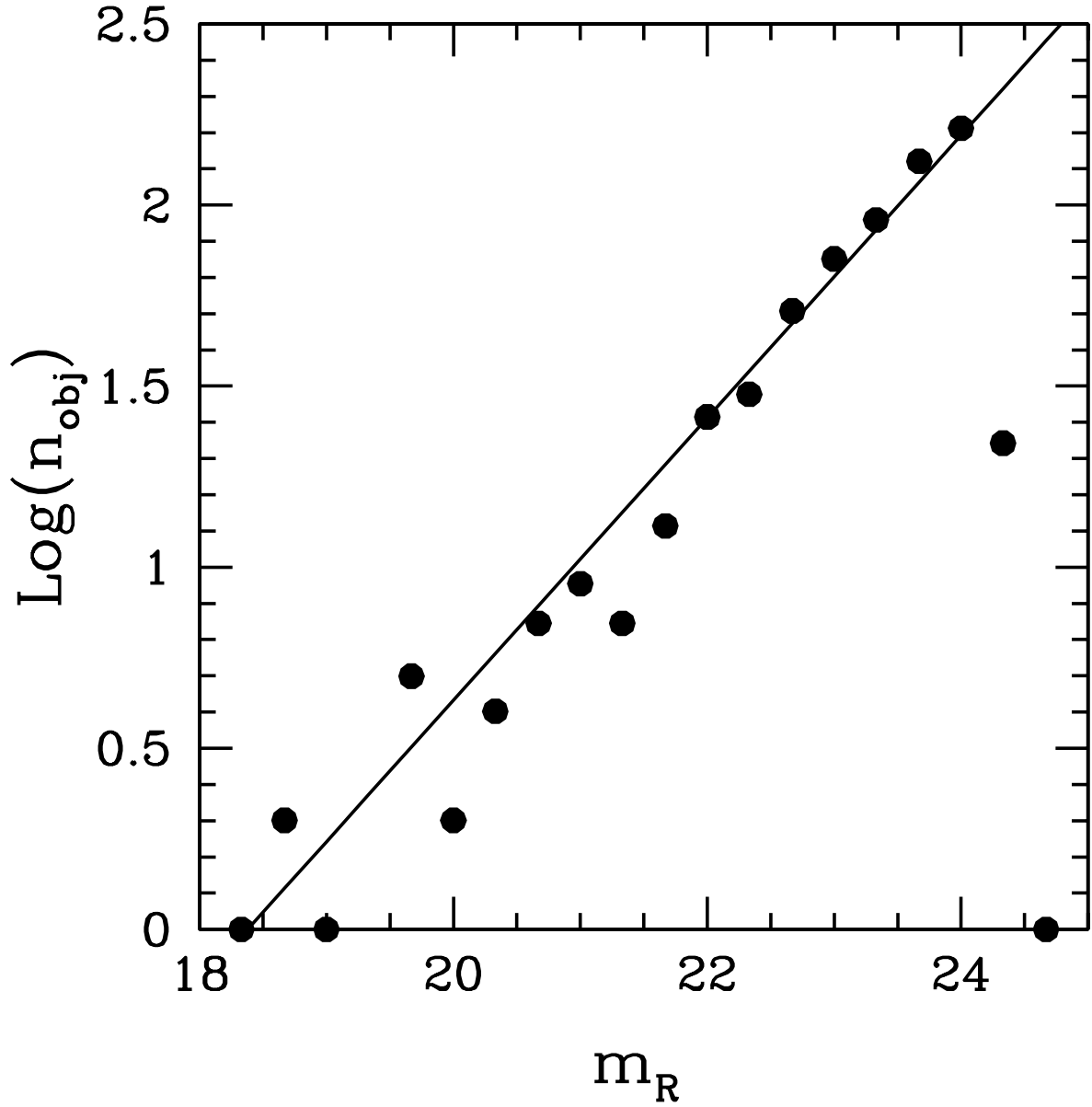}
\caption{Luminosity function of all detected 
objects in the NGC 4874 image  binned to 0.3 mag. Solid 
line represents the BGLF obtained by \citet{T88} and 
scaled to our counts. \label{ngc4874_bglf}}
\end{figure}

Once the residual image of NGC 4874 is created, cosmic-ray events and 
bad pixels are masked. In order to measure $\sigma_{\rm BG}^2$, we
 must detect all objects brighter than $4\sigma$ above the zero mean
 SBF. Using the DAOPHOT task DAOFIND, 649 objects were detected. The 
photometry was made using ALLSTAR. In Fig. \ref{ngc4874_bglf}, the 
luminosity function of all detected objects is shown. The solid line 
represents the BGLF (eq. \ref{bglf}) with $\gamma=0.39$ \citep{T88} 
and scaled to our counts. The BGLF in the case of NGC 4874 is:

\begin{equation} \label{bglfngc4874}
n_{\rm BG}(m)=3.17 \times 10^{-13}  10^{0.39 m} \left({\frac{\rm 
galaxies}{\rm mag \times pix}}\right).
\end{equation}

The fit is very good, so we can conclude that almost all point sources are 
background galaxies, as expected. 

Now, $\sigma_{\rm BG}^2$ can be estimated using this BGLF and eq. \ref{sigma_bg}. The uncertainty in the background variance must be evaluated taking into account the uncertainty of the BGLF slope as well as possible departures from a pure power-law model. Recent determinations of the $R$-band faint galaxy luminosity function result in slopes in the range $\gamma=0.30-0.34$ \citep{SH93}. Considering the result by \citet{T88}, $\gamma=0.39$, we have assumed that $\gamma=0.30-0.40$ is a reasonable range for the slope value. We have done simulations using these two extreme values, resulting in a 20\% variation in the background variance. We assume that this 20\% is the uncertainty in the background variance. In particular, in the case of NGC 4874, for which $m_1^*=24.44$ (see Table \ref{calibr2}) and $m_{\rm c}=23.5$, we obtain: 

\begin{equation} \label{sigma_bg_ngc4874}
\sigma_{\rm BG}^2=(29.4 \pm 5.9)10^{-4} \left(\frac{e^-}{\rm s 
\times pix}\right)^2.
\end{equation}

\begin{figure} 
\plotone{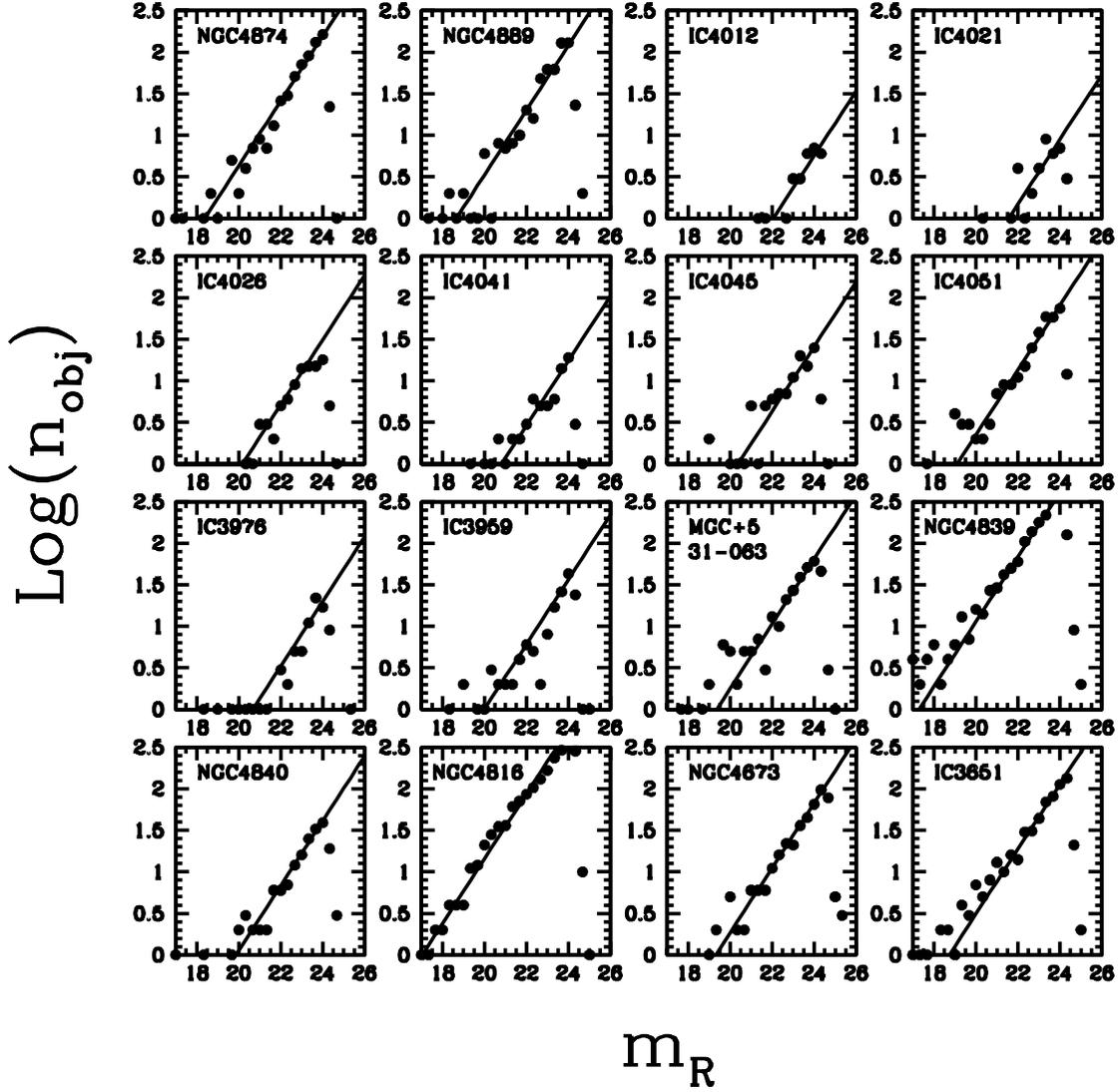}
\caption{Luminosity functions (binned to 0.3 mag) 
of all detected objects in each galaxy studied. Solid lines 
represents the BGLF obtained by \citet{T88} and scaled to
 our counts. \label{bglf.grafi}}
\end{figure}

This procedure is repeated for each galaxy. In Fig. \ref{bglf.grafi}, 
the luminosity functions of all detected objects are presented for each 
galaxy. Again, the solid line represent the BGLF with $\gamma=0.39$, 
scaled to our counts. Using these BGLFs, $\sigma_{\rm BG}^2$ are obtained 
for each galaxy. Results are listed in Table 
\ref{calibr_2}.

\subsection{Measuring \boldmath $\sigma^2_{\rm GC}$\unboldmath and 
the GCS properties}

As in \S \ref{sigmaBG}, we first describe in detail the procedure 
used to measure $\sigma_{\rm GC}^2$ for NGC 4874 and give  the results 
for the  remaining galaxies.

Before computing the power spectrum, all objects brighter than 
$m_{\rm c}=23.5$ are masked and the ring-shaped regions are created 
in the galaxy for the analysis of the spatial distribution of GCs. 
The characteristics of all 
the regions are given in Table \ref{Results} (columns 1 to 3). 
$r_{\rm min}$ and $r_{\rm max}$ are the minimum and  
maximum distances of the selected ring-shaped region to the 
galaxy center, and $A$ is the area of this 
region.

\begin{figure} 
\plotone{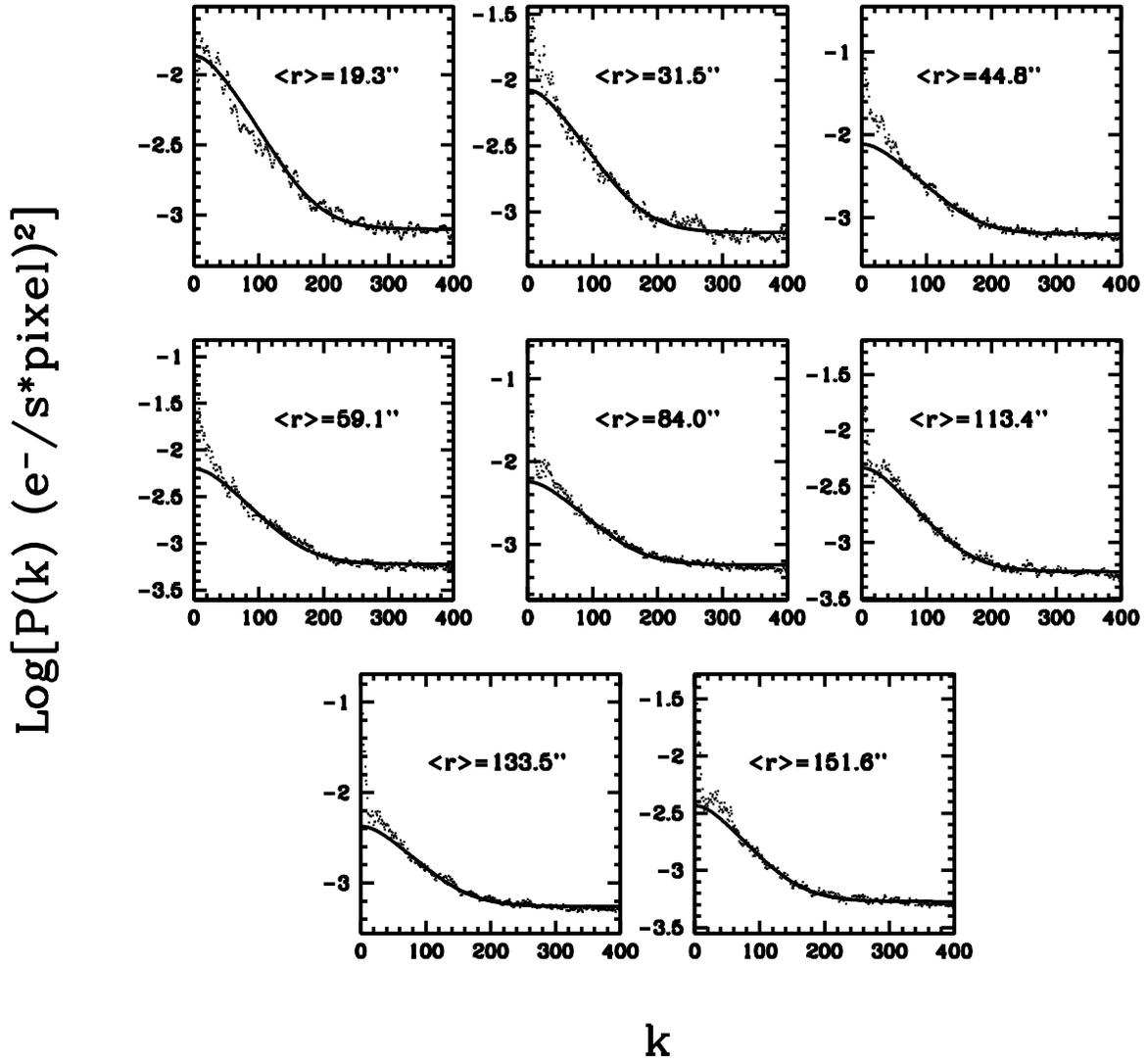}
\caption{SBF analysis of the eight ring-shaped 
regions studied in NGC 4874. Points are the power spectrum 
of the masked ring-shaped region, and solid lines are the 
results of the SBF fit. See details in 
Table \ref{Results}. \label{ngc4874_sbf}}
\end{figure}

The power spectrum of all the masked regions is then obtained 
and the SBF evaluated. In Fig. \ref{ngc4874_sbf}, the SBF fit is 
shown for all the eight ring-shaped regions studied in NGC 4874. 

Spatial variations of the PSF along the CCD mosaic may introduce a significant uncertainty in $P_0$. To limit this efect, we have used a PSF template for each of the four chips of the WFC. Moreover, in order to evaluate the uncertainty produced along a single chip, SBF fit was repeated for NGC 4874 using eleven PSF templates, which were computed using eleven PSF stars located at different positions of the chip.
The diferences in $P_0$ from fit to fit were less than 2\%. This 2\% was added in quadrature 
to the sigma of the $P_0$ measurements of all galaxies. The results for $P_0$ are listed in column 4 of Table \ref{Results}.  

Subtracting $\sigma_{\rm BG}^2$ from $P_0$ in all regions, 
$\sigma_{\rm GC}^2$ is obtained. The results for $\sigma_{\rm GC}^2$ 
are also given in Table \ref{Results} (column 5). Note that NGC 4886 is 
located in the NGC 4889 ring-shaped region with 
$[r_{min},r_{max}]=[71.9 \arcsec, 104.7 \arcsec]$, so $\sigma_{\rm GC}^2$
of NGC 4889 in this region has to be also subtracted to the NGC 4886 $P_0$ results.

The GCLF parameters must be introduced in eq. \ref{sigma-gc} to obtain the number of GCs from $\sigma_{\rm GC}^2$. The GCLF for giant ellipticals seems to be universal. There is no empirical evidence showing galaxy-to-galaxy differences in GCLF parameters \citep{H98}. Moreover, \citet{K99} studied 1057 GCs in the inner region of M87 using {\it HST} observations. This work shows no evidence for radial variations in the GCLF. \citet{K00} found $m_V^0= 27.88 \pm 0.12$ for NGC 4874. This value comes from a constrained fit with $\sigma=1.40$. Assuming a mean value $(R-I) = 0.55$ and $(V-I)= 1.01$ for GCs (Ajhar, Blakeslee, \& Tonry 1994), $m_R^0= 27.42 \pm 0.12$.  

\citet{K00}, based on the reviews of \citet{W96} and \citet{H99}, quote estimates of the intrinsic galaxy-to-galaxy scatters in $\sigma$ and $m_R^0$ of 0.05 and 0.15 mag, respectively. Althought, from \citet{F00}, one might estimate uncertainties twice as large.

In order to include the influence of galaxy-to-galaxy variations in both $\sigma$ and $m_R^0$ in our results, we adopted $\sigma=1.40 \pm 0.05$ and $m_R^0= 27.42 \pm 0.20$ as our fiducial values and tested the influence that 1-sigma variations in both $\sigma$ and $m_R^0$ have on $\sigma^2_{\rm GC}/N_{\rm GC}$. To this purpose, eq. \ref{sigma-gc} was used to evaluate $\sigma^2_{\rm GC}$ using as input parameters the following sets of ($\sigma$, $m_R^0$): (27.22, 1.40), (27.62, 1.40), (24.42, 1.35) and (27.42, 1.45). The adopted results for $\sigma^2_{\rm GC}/N_{\rm GC}$ will be the $mean \pm \sigma$ of the sets' results. 

For NGC 4874 the next relation is obtained:

\begin{equation} \label{ngc4874_signa-gc}
\sigma^2_{\rm GC}= (0.067 \pm 0.016) \times N_{\rm GC}  
\left(\frac{e^-}{\rm s \times pix}\right)^2.
\end{equation}

This calibration depends on the photometric calibration of each target 
field, so it is different for each galaxy. All $\sigma^2_{\rm GC}/N_{\rm GC}$ calibrations 
are presented in Table \ref{calibr_3}, where the ratio 
$\sigma^2_{\rm GC} / N_{\rm GC} $ is given  for the different sets of imput 
parameters (columns 2-5) and for the final adopted value (column 6).

Coming back to NGC 4874, eq. \ref{ngc4874_signa-gc} can be used to 
obtain $N_{\rm GC}$ in each region, and multiplying by the area 
of the selected region, the number $N_{\rm GC}^{\rm region}$ of GCs can be derived 
(see Table \ref{Results}, column 7). If we consider the whole galaxy, the 
total population of GCs is obtained. For NGC 4874, we obtain $17600 \pm 
4100$ GCs. This is the  number of GCs in the whole galaxy except the
 central region, which could not be analyzed due to the difficulty of 
making a good galaxy profile fit near the center. If the  total population 
of GCs is sought, an estimate of the number of GCs in the central 
region is required. In this work, we assume that the surface density of
 GCs in the central region is the same as that obtained in the inner 
ring-shaped region analyzed in the galaxy. This approximation produces 
an underestimate of the real number of GCs in the center of the galaxy, 
but as the size of the  region 
not studied is very small (in some cases negligible), the error introduced 
is also small. In most galaxies, the estimated number of GCs in the 
central region is about 5--10\% of the total population.  

Extrapolating to the central region of the galaxy, we obtain a total 
population of $18200 \pm 4100$ GCs for NGC 4874. 

\begin{figure}
\plotone{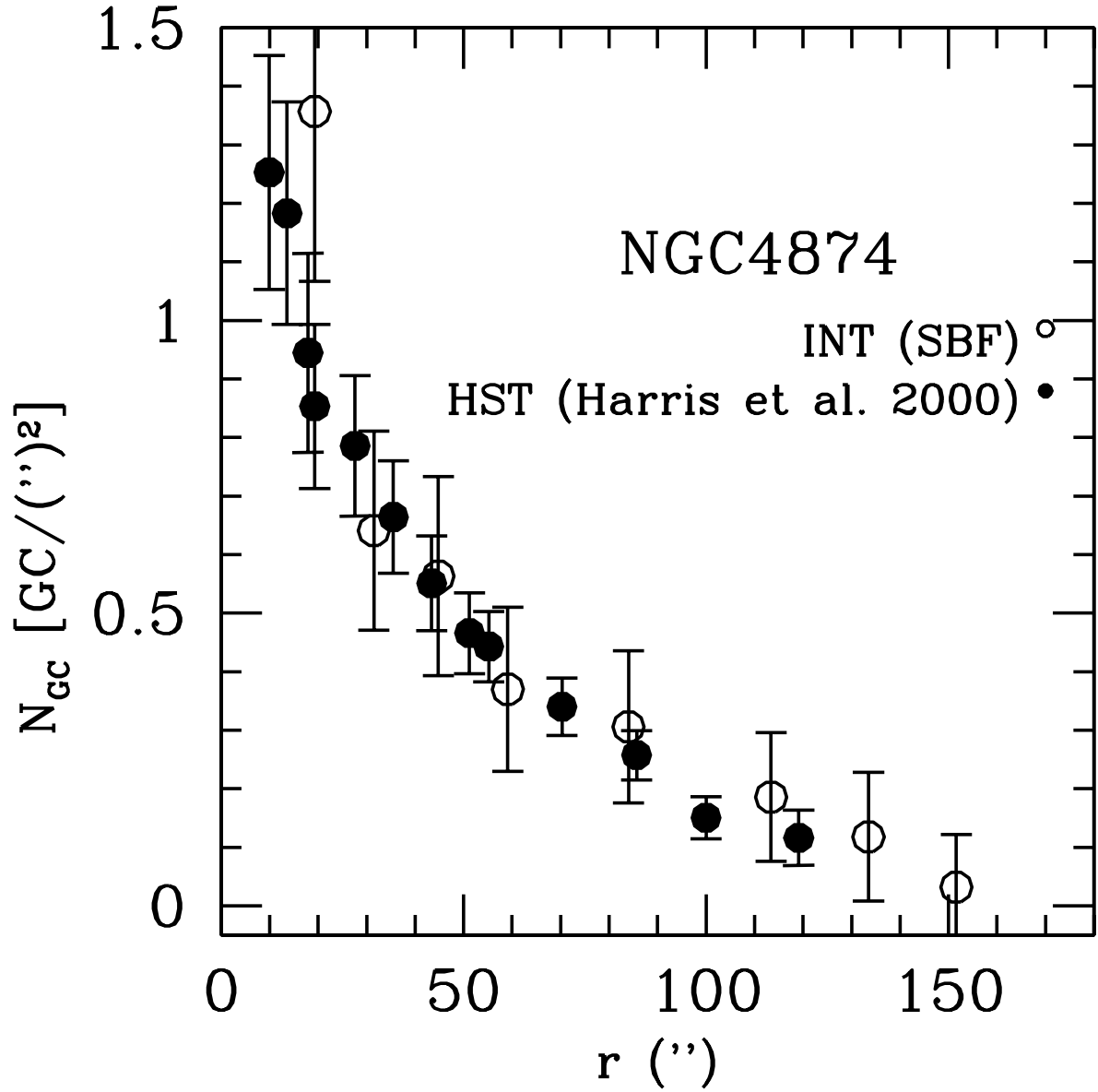}
\caption{Radial dependence of the GCs surface 
density in NGC 4874. Open dots: results obtained with SBF 
in this work; Filled dots: results obtained using the HST 
and traditional techniques \citep{H00}. \label{ngc4874_resultados}}
\end{figure}

In Fig. \ref{ngc4874_resultados}, the radial dependence of the GC surface density is presented. Open dots represent our results obtained with the SBF technique, and filled dots represent the results obtained using the {\it HST} and traditional techniques \citep{H00}. Interestingly, when analyzed with the SBF technique, ground-based data produce similar results to those of the {\it HST} except for the galactic halo, where the surface density of GCs obtained in this work is bigger than that obtained by \citet{H00}. In particular, the total population obtained by \citet{H00} is $9200 \pm 1500$ GCs. The result of the present study is closer to that in \citet{BT95}, who obtained a total number of GCs of $17260 \pm 2030$.

The total number of GCs can be used to derive the specific frequency, $S_N$. The RC3 \citep{RC3} catalogue gives $V^t=11.68 \pm 0.11$ for NGC 4874. In order to transform $V^t$ into $M_V^{\rm TOT}$, some estimation of the Coma cluster distance is required. \citet{Kel00} reported the Cepheid distance to Virgo $(m-M)=31.03 \pm 0.16$. \citet{Jen01} measured the SBF distance of the Virgo galaxies NGC 4472 and NGC 4406, obtaining $(m-M)=31.06 \pm 0.10$ and $(m-M)=31.17 \pm 0.14$ respectively. We adopt the mean of the three previous values as our Virgo distance, i. e.  $(m-M)_{\rm Virgo}=31.08 \pm 0.14$. \citet{K00} using the GCLF obtained a relative Coma-Virgo distance modulus of $4.06 \pm 0.11$. On the other hand, \citet{D97}, reported a fundamental plane relative Coma-Virgo distance of $3.55 \pm 0.15$. Adopting the mean of both, $3.81 \pm 0.16$, the Coma distance results $(m-M)_{\rm Coma}=34.89 \pm 0.20$.

The Coma cluster is located close to the Galactic Pole, with nearly negligible foreground extinction. Using the expresions in \citet{PL95} for the extintion and the K-correction, and asuming $E(B-V)=0.01$  \citep{Ba97} and a redshift $z=0.024$, we obtain $A_R=0.024$ and $K_R=0.025$ mag. After the correction of these effects, we obtain $M_V^{\rm TOT}=-23.26 \pm 0.11$ and hence a specific frequency $S_N=9.0 \pm 2.8$. 

\begin{figure} 
\plotone{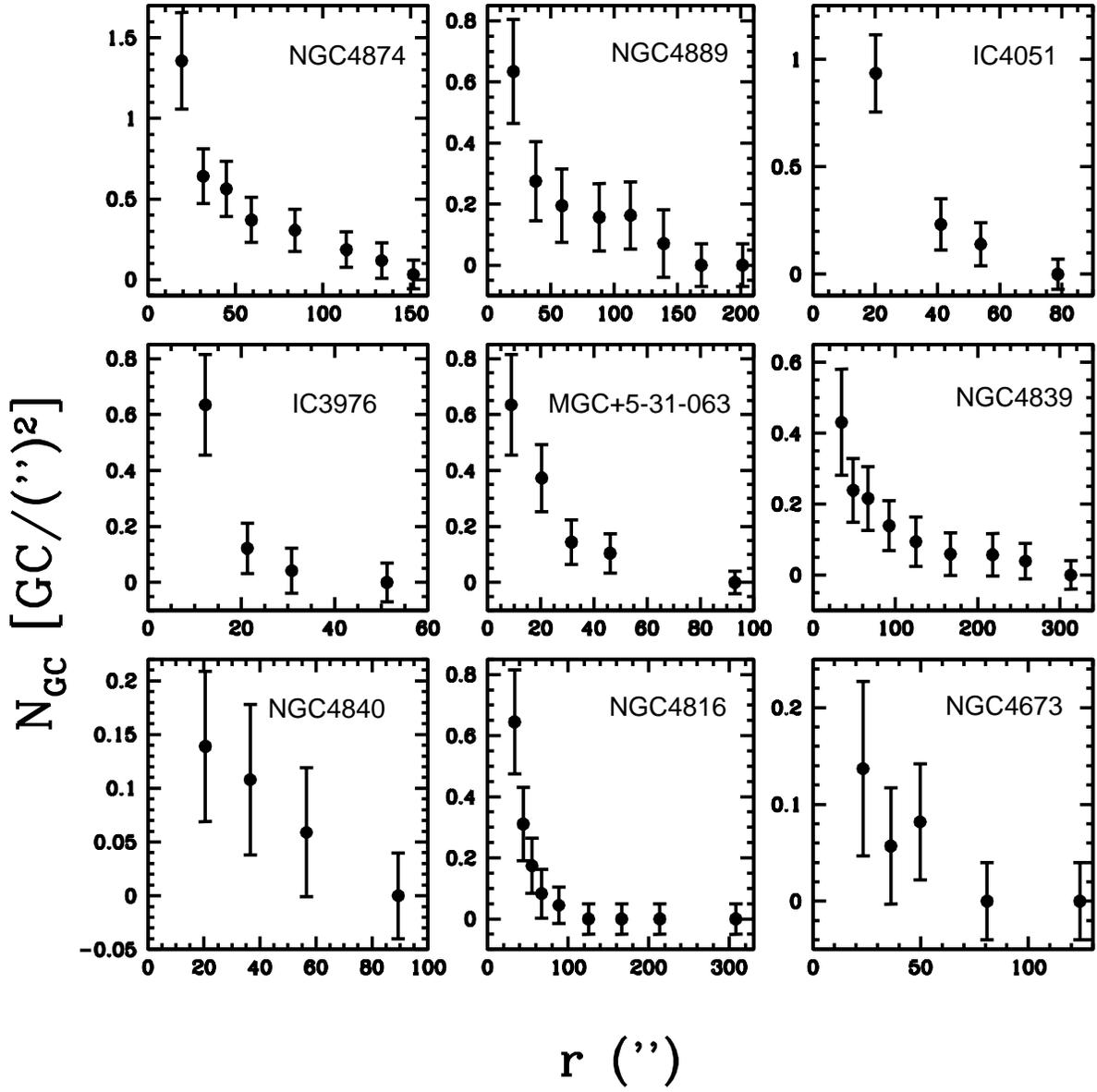}
\caption{Radial structure of the GCS of the nine 
galaxies with high enough angular size. \label{radiales.grafi}}
\end{figure}

For the remaining galaxies, SBF analysis results are presented in Table \ref{Results}, in the same way as for NGC 4874, and for those galaxies that have high enough angular size, the radial structure of those GCSs are shown in Figure \ref{radiales.grafi}. The extrapolation to the central regions of all the galaxies is also shown in the first row of each galaxy in Table \ref{Results}. In Table \ref{sn}, $M_V^{\rm TOT}$ is listed in column 2. For the galaxies IC 3959, MCG +5 $-$31 $-$063, NGC 4673, and IC 3651, $V^t$ does not appear in the RC3 calatogue, and $M_V^{\rm TOT}$ was estimated using $B^t$ and assuming a mean color for these galaxies of $(B-V)=1.1$. In column 3, the distance of the galaxy to the central galaxy NGC 4874 is given, and final results for total populations ($N_{\rm GC}^{\rm tot}$) are listed in column 4 and the derived $S_N$ in column 5. 

The distance uncertainty produces an additional error of 18\% in the results of $S_N$. It must be considered when quoting $S_N$ for a single galaxy. However, it has the same effect in $S_N$ for all the Coma galaxies and does not introduce further internal dispersion in any relative distribution of $S_N$ values of Coma galaxies. In Table \ref{sn}, the uncertainty in the distance has not been included in the $S_N$ results.

\section{Discussion and Conclusions} \label{conclusions}

In the following, we discuss first the scenarios for 
elliptical-galaxy formation in the light of the obtained 
results, and second, the possibility that a merging process is at work in Coma.

\subsection{Elliptical Galaxy Formation Scenarios}

We have measured $S_N$ in 17 elliptical galaxies located in Coma. 
In Table \ref{sn} we present a summary of the obtained results. 
For each galaxy we show the value of $M_V^{\rm TOT}$ (obtained 
from the RC3 value and adopting a distance modulus for Coma of 
$34.89 \pm 0.20$), the distance to the central galaxy,
 NGC 4874, the obtained total population of GCs, and the result 
for $S_N$. The uncertainty in the distance has not been included in the results, it produces an additional error of 18\% in $S_N$.
These results reveal enormous differences in $S_N$ among similar 
galaxies. In particular, $S_N$ varies by an order of magnitude 
from galaxy to galaxy. Extreme cases are: a) at the lower end, 
NGC 4673 has $S_N = 1.0 \pm 0.4$, this value being typical of 
spiral or irregular galaxies, but surprising for an elliptical 
galaxy; b) on the other hand, MCG +5 $-$31 $-$063 has 
$S_N = 13.0 \pm 4.2$ and IC4051 has $S_N = 12.7 \pm 3.2$, which 
are similar to the values found in 
supergiant cD galaxies, but not in ``normal" elliptical galaxies. 

\citet{BTM97} also performed an SBF analysis and reported $S_N$ results
for NGC 4874, NGC 4889 and NGC 4839. They obtained 
$S_N=9.3 \pm 2.0$ for NGC4874; $5.7 \pm 1.3$ for NGC 4889, and 
$4.6 \pm 1.5$ for NGC 4839. These
results are compatible (within the error bars) with ours. We obtained for these galaxies $S_N=9.0 \pm 2.2$, $S_N=4.0 \pm 1.2$ and $S_N=7.0 \pm 1.9$ respectively. Perhaps there is a small difference in the case of NGC 4839. Note that results reported by \citet{BTM97} are ``$metric-S_N$"; i.e., $S_N$ calculated within a radius of 40 kpc. If GCs and halo light follow the same radial distribution, $S_N$ does not change with radius and the ``$metric-S_N$" would be identical to the global $S_N$. The  results for NGC 4839 probably indicate that NGC 4839 GCS is more extended than its halo. Indeed, NGC 4839 has the most extended GCS of our sample (see Fig. \ref{radiales.grafi}).

\citet{WH00} studied {\it HST} images of IC 4051 and proved that a
 central location in a rich cluster environment is not required to
 form a high population of GCs. They obtained a $S_N$ equal to $11 \pm 2$. 
In our work, IC 4051, a ``normal" 
elliptical near the cluster core, has a high $S_N$ of 12.7 
$\pm$ 3.2. Furthermore,  NGC 4874, the central supergiant cD galaxy 
in Coma, also exhibits a relatively high specific frequency, 
$S_N = 9.0 \pm 2.2$. The remaining galaxies studied 
have $S_N$ in the range [2, 7], the mean value being $S_N = 5.1$.

\begin{figure} 
\plotone{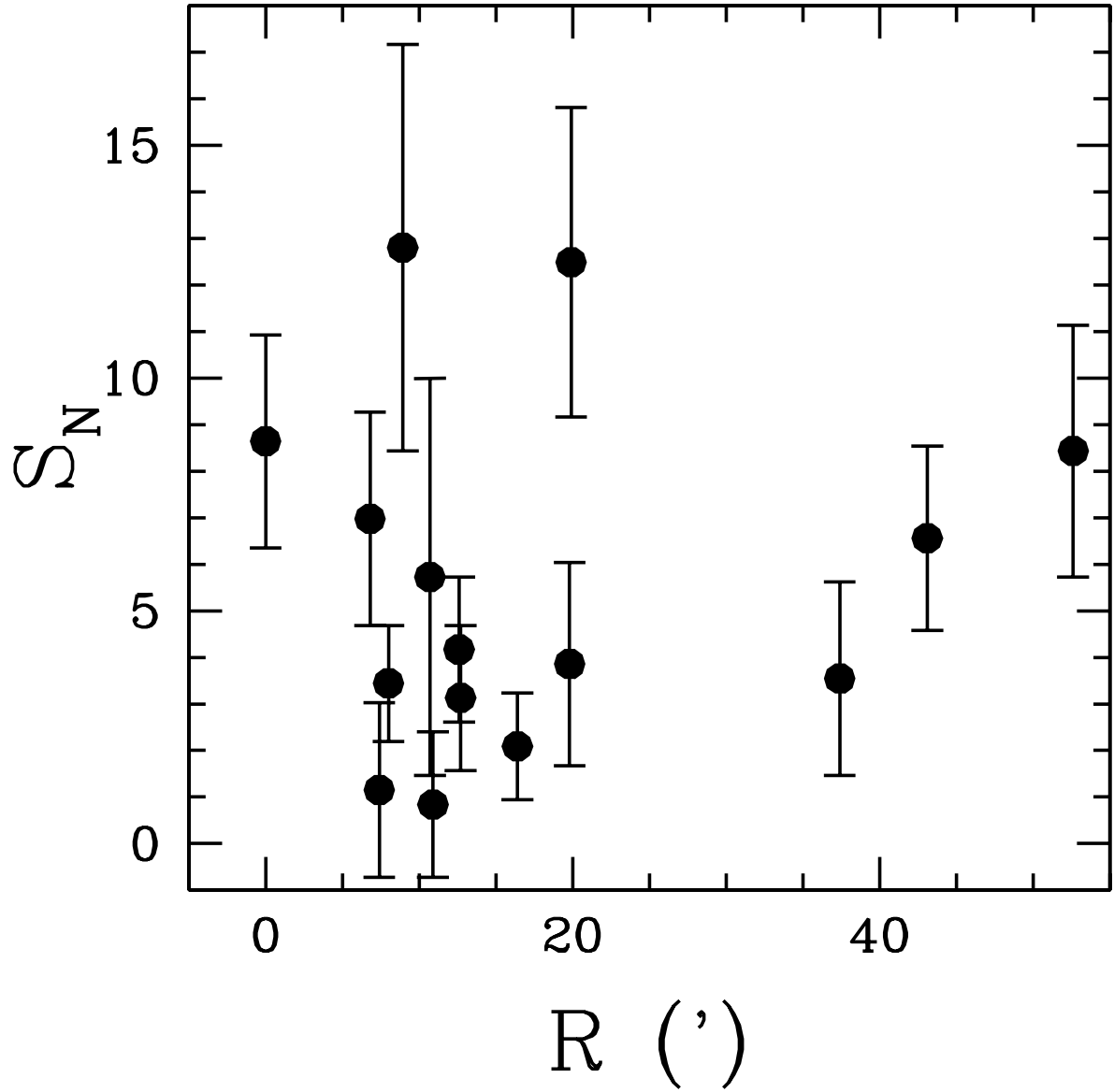}
\caption{$S_N$ versus the distance of the host galaxy
 to the Coma cluster center. \label{snr}}
\end{figure}

\begin{figure}
\plotone{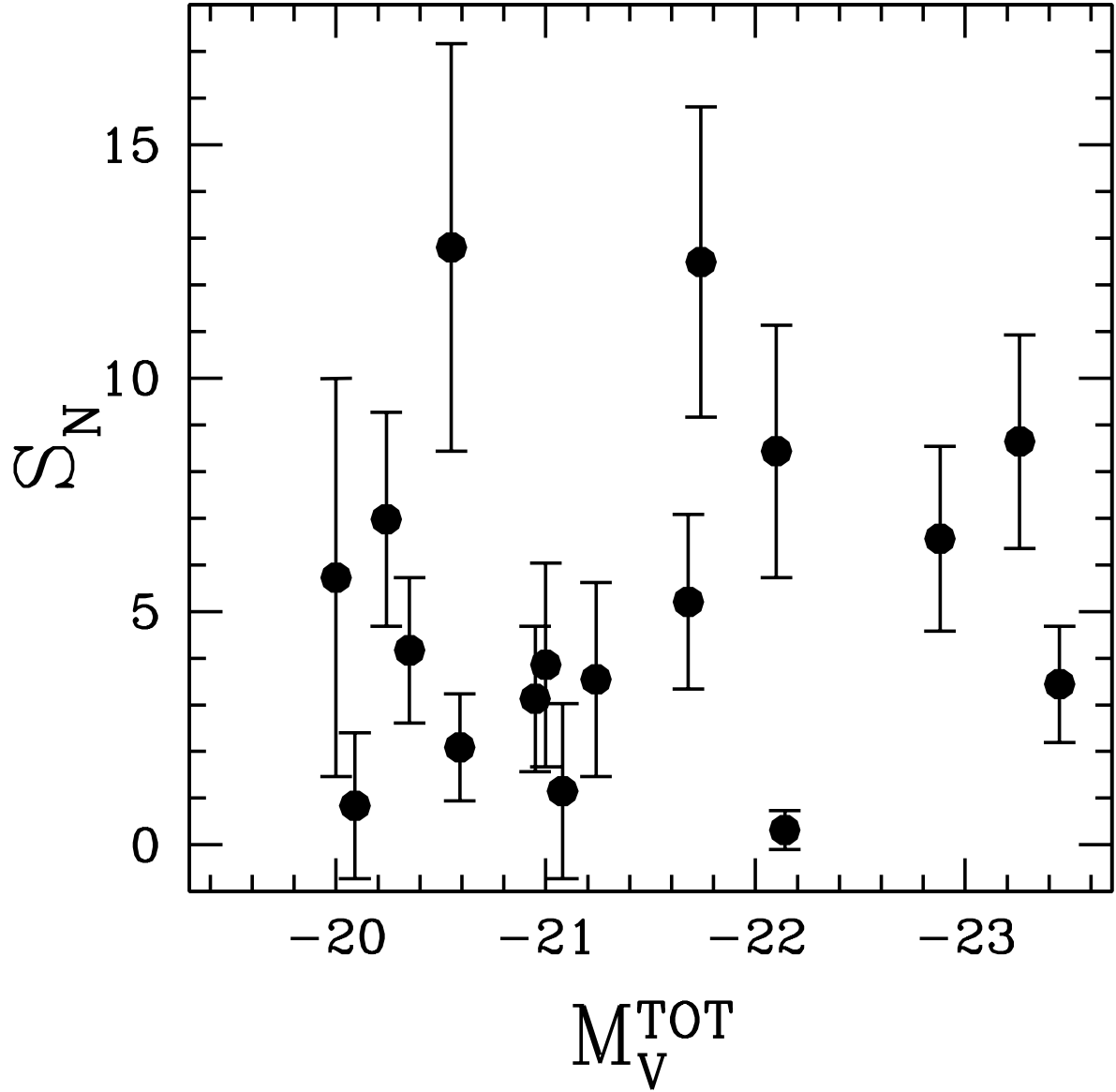}
\caption{$S_N$ versus $M_V^{\rm TOT}$ of the host 
galaxy. \label{snm}}
\end{figure}

Why do IC 4051 and NGC 4673, galaxies with similar absolute 
magnitudes, have differences of a factor of twelve in $S_N$? 
Perhaps this is due to differences in the environment. But 
then, why does
MCG +5 $-$31 $-$063 have $S_N$ five times bigger than IC 4041, 
if both galaxies are located on the border of the Coma cluster 
core and have the same absolute magnitudes?. In order to study possible relations between $S_N$ and 
environment, $S_N$ versus the distance $R$ to the central
 Coma galaxy, NGC 4874 is plotted in Fig. \ref{snr}. No clear trend is found in this plot, which suggests that $S_N$ does not depend 
significantly on the environment in Coma. On the other hand, Fig. \ref{snm} shows $S_N$ versus $M_V^{\rm TOT}$ of each 
galaxy. No relation is found between $S_N$ and
 $M_V^{\rm TOT}$. The figure is completely dominated by 
the dispersion of the points, which is greater than the error 
bars and must therefore be real. 

Formation scenarios for giant ellipticals tend to fall into 
three basic 
classes of models: a) in situ models,  in which
the galaxy condenses by dissipative collapse of gas clouds in 
one or more major bursts, b) mergers of gas-rich systems, 
probably disk-type galaxies, and c) accretion of smaller 
satellites. Various combinations of these extremes are also possible.
 
In situ models predict a correlation between galaxy properties and the 
GCS. If we assume a universal efficiency of GC formation per unit  
total initial galaxy mass \citep{Mc99},

\begin{equation}
\epsilon = \frac{M_{\rm GC}}{M_* + M_{\rm gas}}
\end{equation}
where $\epsilon$ is the efficiency parameter, $M_*$ is the mass 
of the visible stellar component of the galaxy, $M_{\rm gas}$ is
the mass of gas in or around the galactic halo, and $M_{\rm GC}$ is 
the mass of the GCS, then $S_N$ must be related with the luminosity 
of the host galaxy according to:

\begin{equation}
S_N \sim \epsilon (1+\frac{M_{\rm gas}}{M_*})L_{\rm gal}^{0.3}.
\end{equation}

The term $L_{\rm gal}^{0.3}$ accounts for the systematic increase 
in mass-to-light ratio with galaxy luminosity. \citet{Mc99} predicts 
the general trend of $S_N$ with luminosity. The only unknown parameter affecting $S_N$ would be the ratio of gas to stellar mass. In this paper no
evidence is found for a relation between galactic luminosity and $S_N$.

On the other hand, if a range of possible GC formation efficiencies 
is allowed,  the merger model can account for the specific frequency
 range observed for elliptical galaxies \citep{AZ92}. If the merger 
model is to explain the $high-S_N$ phenomenom, GC formation must be 
more efficient in the $high-S_N$ galaxies. 

\citet{F82} and \citet{M87} have suggested that in clusters of galaxies a 
population of intergalactic GCs (IGCs) should exist. A galaxy could 
enlarge its GC population via gravitational capture of IGCs. If they 
are distributed according to the gravitational potential of the whole 
galaxy cluster, a relation between $S_N$ and environment is 
predicted \citep{W87}. The current data set shows no indication supporting this prediction.

The fact that no single scenario seems to account for the observed specific
 frequencies, indicates the history of each galaxy should be deduced 
individually by suitably combining the models mentioned above 
\citep{WH00}. To this aim, it becomes necessary to extend the 
observational information on each galaxy. Besides the specific
 frequency and the radial distribution of the GCS, it is necessary to
 know the details of the subpopulations (when these exist) and the
 kinematics of the GCS. In this way, detailed {\it HST} observations 
of the extreme cases mentioned above (NGC 4673 and MCG +5 $-$31 $-$063)
 could be a good starting point because these peculiar galaxies can
 show characteristics found nowhere else, and so
offer valuable information in testing the different scenarios.

\subsection{Subgroups and Merging in Coma}

\citet{GM01} have recently discovered the existence of three subgroups 
of galaxies in Coma, one of them associated with the cD galaxy NGC 4874 
and the
other two with NGC 4889 and NGC 4839. They conclude that the non-stationarity of the
 dynamical processes at work in the Coma core is due to the merging 
of small-scale groups of galaxies. In this context, each subgroup 
formed separately and then the merger between the different groups 
took place. If this scenario is valid, is there any relation between
 $S_N$ and environment and/or $M_V^{\rm TOT}$ inside each subgroup?

\begin{figure} 
\plotone{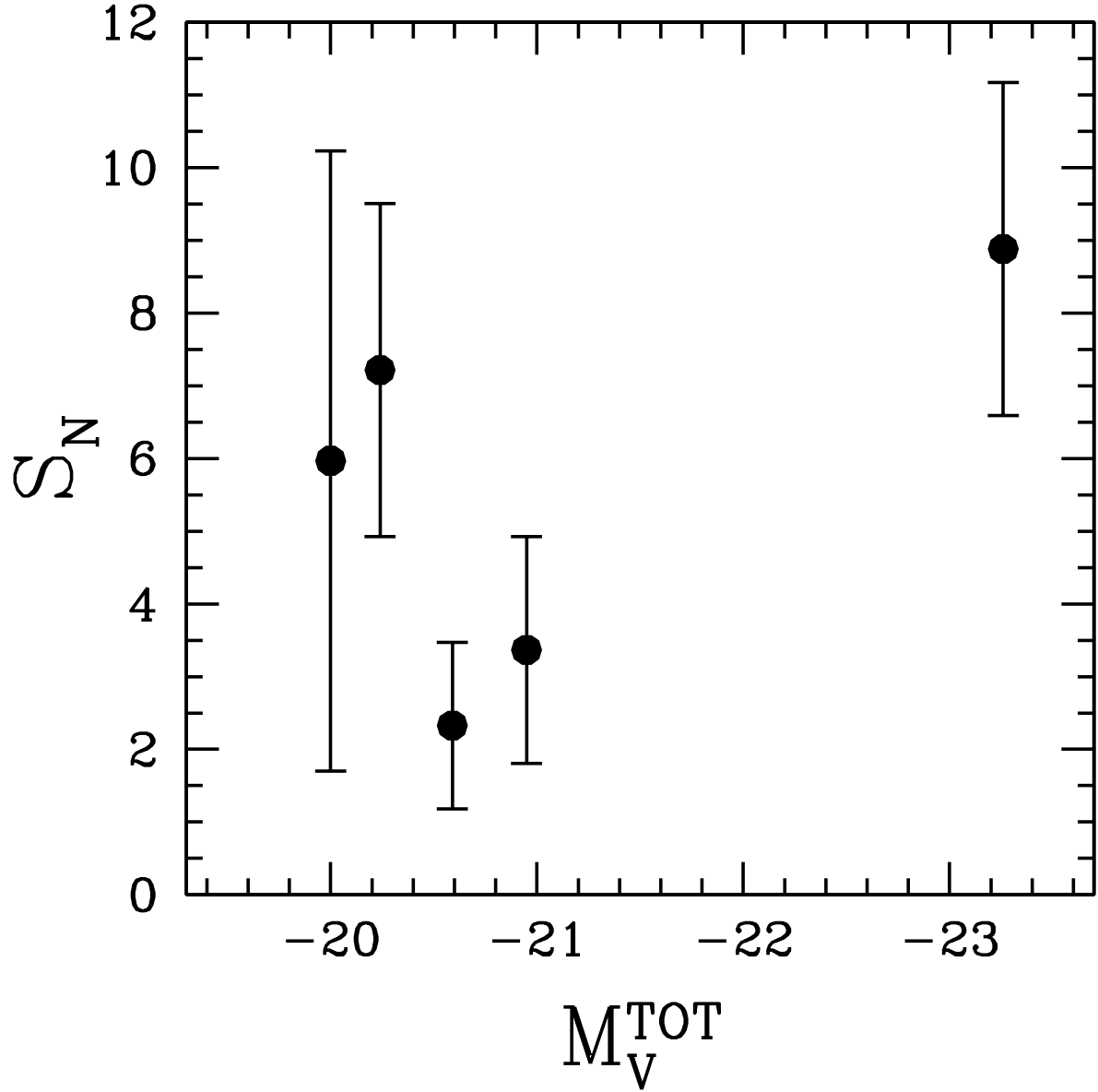}
\caption{$S_N$ versus $M_V^{\rm TOT}$ of the
 host galaxy (for subgroup 2 of Gurzadyan \& Mazure
 2001). \label{snmv2}}
\end{figure}

\begin{figure} 
\plotone{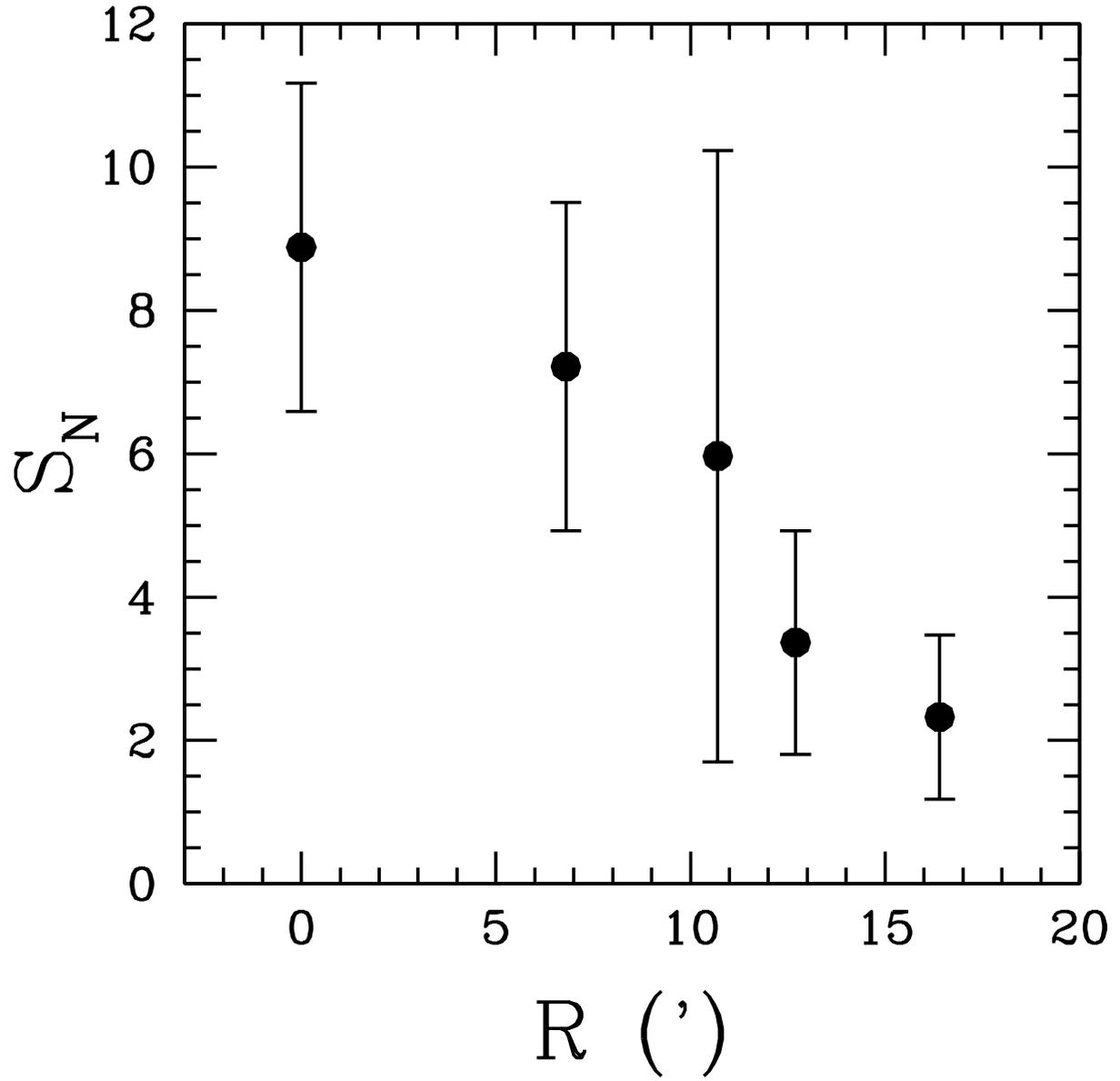}
\caption{$S_N$ versus the distance of the host 
galaxy to the Coma cluster center (for subgroup 2 of
 Gurzadyan \& Mazure 2001). \label{snr2}}
\end{figure}

In order to analyze this question, we restricted our figures to 
galaxies belonging to \citet{GM01} subgroup 2 and studied in this 
paper: NGC 4874, IC 4012, IC 4041, IC 3976, and IC 3959. In 
Fig. \ref{snmv2} $S_N$ versus host galaxy magnitude is plotted, while 
Fig. \ref{snr2} shows $S_N$ versus the distance to the galaxy 
NGC 4874, which is very close to the center of  subgroup 2. No relation between $S_N$ and $M_V^{\rm TOT}$ is found from Fig \ref{snmv2}. But there is an apparent trend in 
Fig. \ref{snr2}: $S_N$ is bigger in high density environments. 
Is this trend real or is it an artifact of the low number of galaxies
 considered?. If this result is confirmed, this will be a strong argument in favor of the IGCs model, so the next step in this work will be
 to enlarge the number of galaxies of the study.

\acknowledgments
 
We are very grateful to W. E. Harris for reading a preliminary version of this paper and to J. J. Fuensalida for his helpful comments.
This article is based on observations made with the 2.5 m Isaac Newton
Telescope operated on the island of La Palma by the ING in the Spanish
Observatorio del Roque de Los Muchachos.
This research has made use of the NASA/IPAC Extragalactic Database (NED) which is operated by the Jet Propulsion Laboratory, California Institute of Technology, under contract with the National Aeronautics and Space Administration. This research has made use of the Digitized Sky Survey,
produced at the Space Telescope Science Institute at Baltimore under U.S.
grant NAGW-2166. This research has been supported by the Instituto de
Astrof\'\i sica de Canarias (grant P3/94), the DGESIC of the Kingdom of
Spain (grant PI97-1438-C02-01), and the DGUI of the autonomous government of
the Canary Islands (grant PI1999/008).

\clearpage

\clearpage

\clearpage

\clearpage

\clearpage

\clearpage

\clearpage

\clearpage

\clearpage

\clearpage

\clearpage

\clearpage

\clearpage

\begin{deluxetable}{cccl}
\tabletypesize{\scriptsize}
\tablecaption{Target fields \label{coords}}
\tablewidth{0pt}
\tablehead{
\colhead{Target field} & \colhead{$\alpha$ (2000)} & 
\colhead{$\delta$(2000)} &
\colhead{Galaxies studied} 
}
\startdata

RC-1 & $13^{\rm h}00^{\rm m}14^{\rm s}.4$ & $+28^{\rm 
o}00\arcmin46\farcs9$ & NGC 4874, NGC 4889, NGC 4886, \\
& & & IC 4012, IC 4021, IC 4026, \\
& & & IC 4041, IC 4045, I C4051, \\
& & & IC 3976, IC 3959, MCG +5 $-$31 $-$063 \\
RC-2 & $12^{\rm h}57^{\rm m}23^{\rm s}.0$ & $+27^{\rm 
o}43\arcmin37\farcs9$ & NGC 4839, NGC 4840, NGC 4816 \\
RC-8 & $13^{\rm h}45^{\rm m}26^{\rm s}.0$ & $+27^{\rm 
o}06\arcmin12\farcs0$ & NGC 4673 \\
RC-9 & $12^{\rm h}40^{\rm m}53^{\rm s}.1$ & $+26^{\rm 
o}43\arcmin39\farcs9$ & IC 3651 \\

\enddata

\end{deluxetable}

\clearpage

\begin{deluxetable}{ccccc}
\tabletypesize{\scriptsize}
\tablecaption{Summary of observations \label{obs}}
\tablewidth{0pt}

\tablehead{
\colhead{Target field} & \colhead{Date} & \colhead{Exposure 
(s)} & \colhead{Filter} & \colhead{Seeing} 
}
\startdata

RC-1 & Apr 25 & 6 $\times$ 300 & $R$ (Sloan) & $1\farcs1$ \\
RC-1 & Apr 27 & 7 $\times$ 300 & $R$ (Sloan) & $1\farcs1$ \\
RC-2 & Apr 27 & 12 $\times$ 300 & $R$ (Sloan) & $1\farcs1$ \\
RC-8 & Apr 27 & 12 $\times$ 300 & $R$ (Sloan) & $1\farcs0$ \\
RC-9 & Apr 27 & 12 $\times$ 300 & $R$ (Sloan) & $1\farcs2$ \\

\enddata

\end{deluxetable}
\clearpage

\begin{deluxetable}{cc}
\tabletypesize{\scriptsize}
\tablecaption{Zero points and colour terms of each chip of the CCD \label{calibr1}}
\tablewidth{0pt}
\tablehead{
\colhead{Zero point} & \colhead{Colour term} 
}
\startdata
$a_1$=24.422 & $b_1$=--0.111 \\
$a_2$=24.671 & $b_2$=--0.320 \\
$a_3$=24.682 & $b_3$=--0.178 \\
$a_4$=24.654 & $b_4$=--0.132 \\

\enddata

\end{deluxetable}\clearpage

\begin{deluxetable}{lc}
\tabletypesize{\scriptsize}
\tablecaption{Photometric calibration \label{calibr2}}
\tablewidth{0pt}
\tablehead{
\colhead{Galaxy} &\colhead{$m_1^*$} 
}

\startdata

NGC 4874    & 24.44 $\pm$ 0.01 \\
NGC 4889    & 24.44 $\pm$ 0.01 \\
NGC 4886    & 24.44 $\pm$ 0.01 \\
IC 4012     & 24.44 $\pm$ 0.01 \\
IC 4021   & 24.44 $\pm$ 0.01 \\
IC 4026      & 24.44 $\pm$ 0.01 \\
IC 4041       & 24.44 $\pm$ 0.01 \\
IC 4045      & 24.44 $\pm$ 0.01 \\
IC 4051    & 24.44 $\pm$ 0.01 \\
IC 3976   & 24.21 $\pm$ 0.01 \\
IC 3959     & 24.36 $\pm$ 0.02 \\
MCG +5 --31 --063    &  24.36 $\pm$ 0.02 \\
NGC 4839   & 24.22 $\pm$ 0.01 \\
NGC 4840     & 24.22 $\pm$ 0.01 \\
NGC 4816    & 24.37 $\pm$ 0.02 \\
NGC 4673     & 24.43 $\pm$ 0.01 \\
IC 3651    & 24.43 $\pm$ 0.01 \\

\enddata

\end{deluxetable}

\clearpage

\begin{deluxetable}{crcccc}
\tabletypesize{\scriptsize}
\tablecaption{Results of the SBF consistency 
test  \label{testing}}
\tablewidth{0pt}
\tablehead{
\colhead{Experiment} & \colhead{$\sigma^2_{\rm noise}$ 
($e^-$/pixel)$^2$} & \colhead{$N_{\rm GC}^{\rm in}$}&
\colhead{ $P_1$ ($e^-$/pixel)$^2$} &
\colhead{ $P_0$ ($e^-$/pixel)$^2$} & \colhead{$N_{\rm 
GC}^{\rm out}$}
}

\startdata

(a)  & 0     & 10000 & 2.55  & 2920 $\pm$ 40   & 10500 $\pm$ 140  \\
(b)  & 625   & 10000 & 628   & 2890 $\pm$ 60   & 10400 $\pm$ 200  \\
(c)  & 3025  & 10000 & 3027  & 2900 $\pm$ 100  & 10400 $\pm$ 400  \\
(d)  & 4900  & 10000 & 4900  & 2900 $\pm$ 150  & 10400 $\pm$ 500  \\
(e)  & 10000 & 10000 & 9994  & 3000 $\pm$ 300  & 10800 $\pm$ 1100  \\
(f)  & 15625 & 10000 & 15568 & 3500 $\pm$ 500  & 12600 $\pm$ 1800  \\

\enddata

\end{deluxetable}

\clearpage
\begin{deluxetable}{lc}
\tabletypesize{\scriptsize}
\tablecaption{Results for $\sigma^2_{\rm BG}$, in units of $10^{-4}\times\left(\frac{e^-}{\rm s \times pix}\right)^2$. \label{calibr_2}}
\tablewidth{0pt}
\tablehead{
\colhead{Galaxy} & \colhead{$\sigma^2_{\rm BG}$} 
}

\startdata

NGC 4874  & 29.4 $\pm$ 5.9 \\
NGC 4889  & 26.0 $\pm$ 5.2 \\
NGC 4886  & 26.0 $\pm$ 5.2 \\
IC 4012   & 18.0 $\pm$ 3.6 \\
IC 4021   & 16.0 $\pm$ 3.2 \\
IC 4026   & 17.0 $\pm$ 3.4 \\
IC 4041   & 20.0 $\pm$ 4.0 \\
IC 4045   & 20.0 $\pm$ 4.0 \\
IC 4051   & 26.0 $\pm$ 5.2 \\
IC 3976   & 12.1 $\pm$ 2.4 \\
IC 3959   & 15.3 $\pm$ 3.1 \\
MCG +5 --31 --063 & 13.7 $\pm$ 2.7 \\
NGC 4839  & 10.1 $\pm$ 2.0 \\
NGC 4840  & 10.0 $\pm$ 2.0 \\
NGC 4816  & 17.5 $\pm$ 3.5 \\
NGC 4673  & 14.2 $\pm$ 2.8 \\
IC 3651   & 13.7 $\pm$ 2.7 \\

\enddata

\end{deluxetable}

\clearpage

\begin{deluxetable}{ccccccc}
\tabletypesize{\scriptsize}
\tablecaption{Properties of the ring-shaped regions analyzed in each galaxy and SBF results \label{Results}}

\tablewidth{0pt}
\tablehead{

\colhead{$r_{\rm min}$} & 
\colhead{$r_{\rm max}$} & 
\colhead{$A$} & 
\colhead{$P_0 \times 10^4$} &  
\colhead{$\sigma_{\rm GC}^2 \times 10^4$}  & 
\colhead{$N_{\rm GC}$} & 
\colhead{$N_{\rm GC}^{\rm region}$} \\

$(\arcsec)$ & 
$(\arcsec)$ & 
$(\arcsec)^2$ &  
$\left(\frac{e^-}{\rm s \times pix}\right)^2$ \tablenotemark{(a)} &   
$\left(\frac{e^-}{\rm s \times pix}\right)^2$ \tablenotemark{(a)} & 
${\rm GC}/(\arcsec)^2$ &  } 

\startdata

\multicolumn{7}{c}{\bf {NGC 4874}} \\
\hline
0     & 12.6  & 452   & -                  &  -                   & -     &  613 $\pm$ 130  \tablenotemark{(b)} \\
12.6  & 25.9  & 1410  & 130    $\pm$ 13    &  101    $\pm$ 14     & 1.36  $\pm$ 0.29   & 1910  $\pm$ 410    \\
25.9  & 37.0  & 1479  & 77.0   $\pm$ 3.4   &  47.6   $\pm$ 6.8    & 0.64 $\pm$ 0.29   & 950   $\pm$ 250    \\
37.0  & 52.5  & 3781  & 71.2   $\pm$ 3.3   &  41.8   $\pm$ 6.8    & 0.56 $\pm$ 0.17   & 2130  $\pm$ 640    \\
52.5  & 65.6  & 6408  & 56.9   $\pm$ 3.2   &  27.5   $\pm$ 6.7    & 0.37 $\pm$ 0.17   & 2370  $\pm$ 900    \\
65.6  & 102.3 & 13022 & 52.1   $\pm$ 2.2   &  22.7   $\pm$ 6.3    & 0.31 $\pm$ 0.14   & 4000  $\pm$ 1700   \\
102.3 & 125.1 & 15031 & 42.3   $\pm$ 1.2   &  13.8   $\pm$ 6.0    & 0.19 $\pm$ 0.13   & 2800  $\pm$ 1600    \\
125.1 & 141.8 & 24153 & 38.2   $\pm$ 3.1   &   8.8   $\pm$ 6.7    & 0.12 $\pm$ 0.11   & 2900  $\pm$ 2600    \\
141.8 & 161.4 & 18778 & 31.8   $\pm$ 1.2   &   2.4   $\pm$ 6.0    & 0.03 $\pm$ 0.09   & 600   $\pm$ 1700    \\

\hline
\multicolumn{7}{c}{\bf{NGC 4889}} \\
\hline

0     & 10.4  & 250   & -          &   -        &  -     &  158 $\pm$ 42 \tablenotemark{(b)} \\
10.4  & 30.5  & 3161  & 73.1 $\pm$ 3.3 & 47.1 $\pm$ 6.2 & 0.63 $\pm$ 0.17 & 2000 $\pm$  540  \\
30.5  & 45.7  & 4722  & 46.4 $\pm$ 3.1 & 20.4 $\pm$ 6.1 & 0.27 $\pm$ 0.13 & 1300 $\pm$  610  \\
45.7  & 71.9  & 12767 & 40.5 $\pm$ 3.1 & 14.5 $\pm$ 6.1 & 0.19 $\pm$ 0.12 & 2500 $\pm$  1500  \\
71.9  & 104.7 & 10904 & 37.7 $\pm$ 3.1 & 11.7 $\pm$ 6.1 & 0.16 $\pm$ 0.11 & 1700 $\pm$  1200  \\
104.7 & 121.2 & 6287  & 38.1 $\pm$ 3.1 & 12.1 $\pm$ 6.1 & 0.16 $\pm$ 0.11 & 1020 $\pm$  690  \\
121.2 & 157.1 & 12558 & 31.3 $\pm$ 5.0 &  5.3 $\pm$ 7.2 & 0.07 $\pm$ 0.11 & 900  $\pm$  1400  \\
157.1 & 180.3 & 15044 & 26.9 $\pm$ 3.0 &  0      &  0 & 0  \\
180.3 & 222.3 & 22139 & 26.2 $\pm$ 2.1 &  0      &  0 & 0     \\

\hline
\multicolumn{7}{c}{\bf{NGC 4886}\tablenotemark{(c)}} \\
\hline

0    & 3.5  & 45   &  -             &   -            &  -    & 26  $\pm$ 9 \tablenotemark{(b)} \\
3.5  & 24.1 & 1628 & 80.2 $\pm$ 6.4 & 42.5 $\pm$ 10.2& 0.57  $\pm$ 0.20  & 931  $\pm$  325 \\
24.1 & 33.4 & 2727 & 45.0 $\pm$ 4.1 & 7.3  $\pm$ 8.9 & 0.10  $\pm$ 0.13  & 267  $\pm$  354   \\

\hline
\multicolumn{7}{c}{\bf{IC 4012}} \\
\hline

0    & 6.3   & 127  &  -              &  -                    &   -    & 17 $\pm$ 11  \tablenotemark{(b)} \\
6.3  & 39.0  & 4605 & 27.8 $\pm$ 3.1 & 9.8 $\pm$ 4.8 & 0.13 $\pm$ 0.09 & 610  $\pm$ 410 \\
39.0 & 46.7  & 3692 & 18.0 $\pm$ 1.1 & 0         & 0      &  0    \\

\hline
\multicolumn{7}{c}{\bf{IC 4021}} \\
\hline

0    & 5.0  & 85   &  -             &     -     &  -   &  6  $\pm$ 6  \tablenotemark{(b)} \\
5.0  & 27.1 & 2229 & 21.0 $\pm$ 1.1 & 5.0 $\pm$ 3.4 & 0.07 $\pm$ 0.07 & 150 $\pm$ 160   \\
27.1 & 46.7 & 6011 & 16.4 $\pm$ 1.0 &  0        & 0    & 0    \\

\hline
\multicolumn{7}{c}{\bf{IC 4026}} \\
\hline

0    & 6.5   & 129   & -            &   -         &  -    & 78  $\pm$ 22 \tablenotemark{(b)} \\
6.5  & 16.5  & 694   & 62.1 $\pm$ 6.1 & 45.1 $\pm$ 7.0 & 0.61 $\pm$ 0.17 & 421 $\pm$ 120    \\
16.5 & 28.2  & 1915  & 22.8 $\pm$ 3.1 &  5.8 $\pm$ 4.6 & 0.08 $\pm$ 0.08 & 150 $\pm$ 150   \\
28.2 & 56.7  & 16810 & 17.2 $\pm$ 1.1 & 0           & 0     & 0    \\

\hline
\multicolumn{7}{c}{\bf{IC 4041}} \\
\hline

0    & 6.4   & 249   &  -         &  -         & -     & 74  $\pm$ 30 \tablenotemark{(b)} \\
6.4  & 25.8  & 1336  & 42.2 $\pm$ 3.1 & 22.2 $\pm$ 5.1 & 0.30 $\pm$ 0.12 & 399 $\pm$ 160   \\
25.8 & 44.3  & 11113 & 21.2 $\pm$ 1.1 & 0          & 0     & 0  \\

\hline
\multicolumn{7}{c}{\bf{IC 4045}} \\
\hline

0    & 13.7 & 675   &  -         &    -       & -      & 146 $\pm$ 74 \tablenotemark{(b)} \\
13.7 & 37.3 & 3654  & 36.1 $\pm$ 3.1 & 16.1  $\pm$ 5.1 & 0.22 $\pm$ 0.11 & 790 $\pm$ 400   \\
37.3 & 52.5 & 3841  & 23.0 $\pm$ 3.0 &  3.0  $\pm$ 5.0 & 0.04 $\pm$ 0.08 & 150 $\pm$ 310   \\
52.5 & 74.7 & 13400 & 19.7 $\pm$ 1.1 & 0          & 0      & 0\\

\hline
\multicolumn{7}{c}{\bf{IC 4051}} \\
\hline

0    & 5.5  & 126   &  -         &  -         &   -    &  118  $\pm$ 23 \tablenotemark{(b)} \\
5.5  & 34.9 & 4093  & 95.5 $\pm$ 4.4 & 69.5 $\pm$ 6.8 & 0.94 $\pm$ 0.18 &  3830 $\pm$  730  \\
34.9 & 47.2 & 4367  & 43.2 $\pm$ 3.1 & 17.2 $\pm$ 6.1 & 0.23 $\pm$ 0.12 &  1010 $\pm$  520 \\
47.2 & 60.6 & 9216  & 36.4 $\pm$ 2.1 & 10.4 $\pm$ 5.6 & 0.14 $\pm$ 0.10 &  1290 $\pm$  920  \\ 
60.6 & 96.9 & 23739 & 26.1 $\pm$ 2.1 & 0          &  0     & 0 \\

\hline
\multicolumn{7}{c}{\bf{IC 3976}} \\
\hline

0    & 8.0   & 196  & -          &   -        &   -    & 124 $\pm$ 35 \tablenotemark{(b)} \\
8.0  & 16.7  & 862  & 43.8 $\pm$ 6.1 & 31.7 $\pm$ 6.6 & 0.64  $\pm$ 0.18  & 550 $\pm$ 150  \\
16.7 & 25.9  & 1363 & 18.2 $\pm$ 3.0 & 6.1  $\pm$ 3.8 & 0.12 $\pm$ 0.09 & 170 $\pm$ 120  \\
25.9 & 36.2  & 1900 & 14.2 $\pm$ 3.0 & 2.1  $\pm$ 3.8 & 0.04 $\pm$ 0.08 & 80  $\pm$ 150  \\
36.2 & 66.6  & 6055 & 11.5 $\pm$ 3.0 & 0          &  0     & 0   \\

\hline
\multicolumn{7}{c}{\bf{IC 3959}} \\
\hline

0    & 11.5  & 459   &  -             &  -            &  -    &  105 $\pm$ 46 \tablenotemark{(b)} \\
11.5 & 33.2  & 2878  & 30.2 $\pm$ 2.1 & 14.9 $\pm$ 3.7 & 0.23 $\pm$ 0.10 & 660  $\pm$ 280   \\
33.2 & 44.9  & 2821  & 18.1 $\pm$ 1.1 & 2.8  $\pm$ 3.3 & 0.04 $\pm$ 0.06 & 120  $\pm$ 170    \\
44.9 & 89.8  & 22707 & 15.2 $\pm$ 1.0 & 0             & 0     & 0    \\

\hline
\multicolumn{7}{c}{\bf{MCG +5 $-$31 $-$063}} \\
\hline

0    & 1.9    & 12    & -              &   -            &  -     &  8  $\pm$ 2 \tablenotemark{(b)} \\
1.9  & 15.9   & 765   & 55.3 $\pm$ 7.1  &  41.6  $\pm$ 7.6 & 0.64  $\pm$ 0.18 & 486 $\pm$ 140  \\
15.9 & 24.9   & 1317  & 38.1 $\pm$ 3.1  &  24.4  $\pm$ 4.1 & 0.37 $\pm$ 0.12 & 491 $\pm$ 160  \\
24.9 & 38.3   & 2887  & 23.1 $\pm$ 2.0  &  9.4   $\pm$ 3.4 & 0.14 $\pm$ 0.08 & 420 $\pm$ 230  \\
38.3 & 54.0   & 7195  & 20.5 $\pm$ 0.7  &  6.8   $\pm$ 2.8 & 0.10 $\pm$ 0.07 & 750 $\pm$ 570 \\
54.0 & 133.2  & 43212 & 13.5 $\pm$ 0.6 &  0             & 0      & 0   \\

\hline
\multicolumn{7}{c}{\bf{NGC 4839}} \\
\hline

0     & 31.6   & 3144   & -              &  -             &  -    & 1355 $\pm$ 470 \tablenotemark{(b)} \\
31.6  & 38.2   & 1730   & 32.1 $\pm$ 5.0 & 22.0 $\pm$ 5.4 & 0.43  $\pm$ 0.15  & 750  $\pm$ 260   \\
38.2  & 60.5   & 3880   & 22.3 $\pm$ 1.1 & 12.2 $\pm$ 2.3 & 0.24 $\pm$ 0.09 & 930  $\pm$ 350   \\
60.5  & 73.6   & 6452   & 22.1 $\pm$ 2.1 & 11.0 $\pm$ 2.9 & 0.22 $\pm$ 0.09 & 1390 $\pm$ 580  \\
73.6  & 111.9  & 7756   & 18.2 $\pm$ 1.1 &  7.1 $\pm$ 2.3 & 0.14 $\pm$ 0.07 & 1080 $\pm$ 540   \\
111.9 & 138.5  & 13850  & 14.9 $\pm$ 1.6 &  4.8 $\pm$ 2.6 & 0.09 $\pm$ 0.07 & 1300 $\pm$ 970   \\
138.5 & 197.2  & 13249  & 13.1 $\pm$ 1.0 &  3.0 $\pm$ 2.2 & 0.06 $\pm$ 0.06 & 780  $\pm$ 790  \\
197.2 & 240.1  & 24147  & 13.0 $\pm$ 1.0 &  2.9 $\pm$ 2.2 & 0.06 $\pm$ 0.06 & 1400 $\pm$ 1400   \\
240.1 & 277.4  & 25663  & 12.1 $\pm$ 1.0 &  2.0 $\pm$ 2.2 & 0.04 $\pm$ 0.05 & 1000 $\pm$ 1300   \\
277.4 & 349.6  & 259328 & 10.3 $\pm$ 1.0 & 0              & 0     & 0 \\

\hline
\multicolumn{7}{c}{\bf{NGC 4840}} \\
\hline

0    & 12.9  & 545   & -              &  -            &  -      & 76  $\pm$ 38 \tablenotemark{(b)} \\
12.9 & 28.1  & 1705  & 17.1 $\pm$ 1.6 & 7.1 $\pm$ 2.6 & 0.14 $\pm$ 0.07 & 240  $\pm$ 120 \\
28.1 & 44.9  & 4182  & 15.5 $\pm$ 1.6 & 5.5 $\pm$ 2.6 & 0.11 $\pm$ 0.07 & 450  $\pm$ 290 \\
44.9 & 69.2  & 8811  & 13.0 $\pm$ 0.8 & 3.0 $\pm$ 2.2 & 0.06 $\pm$ 0.06 & 520  $\pm$ 530 \\
69.2 & 108.9 & 18358 & 9.5  $\pm$ 0.8 & 0       &  0      & 0  \\

\hline
\multicolumn{7}{c}{\bf{NGC 4816}} \\
\hline

0     & 28.0  & 2464   & -              &   -            &   -   & 1590 $\pm$ 420 \tablenotemark{(b)} \\
28.0  & 39.5  & 2357   & 60.4 $\pm$ 6.1 & 42.9 $\pm$ 7.0 & 0.65  $\pm$ 0.17  & 1520 $\pm$ 400   \\
39.5  & 49.6  & 3053   & 38.2 $\pm$ 3.1 & 20.7 $\pm$ 4.7 & 0.31  $\pm$ 0.12 & 950  $\pm$ 370  \\
49.6  & 61.3  & 4313   & 29.1 $\pm$ 1.1 & 11.6 $\pm$ 3.7 & 0.17  $\pm$ 0.09 & 750  $\pm$ 390   \\
61.3  & 73.2  & 6025   & 23.0 $\pm$ 2.1 &  5.5 $\pm$ 4.1 & 0.08  $\pm$ 0.08 & 500  $\pm$ 480   \\
73.2  & 104.3 & 18638  & 20.5 $\pm$ 0.7 &  3.0 $\pm$ 3.6 & 0.04  $\pm$ 0.06 & 800  $\pm$ 1100  \\
104.3 & 145.4 & 34964  & 17.2 $\pm$ 0.6 & 0              & 0     & 0    \\
145.4 & 186.1 & 44838  & 17.4 $\pm$ 1.1 & 0              & 0     & 0   \\
186.1 & 241.4 & 45584  & 17.9 $\pm$ 1.1 & 0              & 0     & 0   \\
241.4 & 374.0 & 193496 & 16.5 $\pm$ 1.1 & 0              & 0     & 0   \\

\hline
\multicolumn{7}{c}{\bf{NGC 4673}} \\
\hline

0     & 15.6   & 770   & -              &   -           &  -     & 105 $\pm$ 69 \tablenotemark{(b)} \\
15.6  & 30.8   & 1810  & 24.4 $\pm$ 4.1 & 10.2 $\pm$ 5.0 & 0.14 $\pm$ 0.09 & 250 $\pm$ 160   \\
30.8  & 41.3   & 2370  & 18.4 $\pm$ 0.7 &  4.2 $\pm$ 2.9 & 0.06 $\pm$ 0.06 & 135 $\pm$ 140   \\
41.3  & 58.1   & 3140  & 20.3 $\pm$ 1.1 &  6.1 $\pm$ 3.0 & 0.08 $\pm$ 0.06 & 260 $\pm$ 190  \\
58.1  & 103.5  & 13364 & 14.0 $\pm$ 0.6 & 0             & 0      & 0    \\
103.5 & 144.4  & 36976 & 15.5 $\pm$ 0.6 & 0             & 0      & 0  \\

\hline
\multicolumn{7}{c}{\bf{IC 3651}} \\
\hline

0    & 17.6  & 1005  & 62.9 $\pm$ 6.1 & 49.2 $\pm$ 6.7 & 0.66 $\pm$ 0.17 & 670 $\pm$ 170   \\
17.6 & 39.0  & 3810  & 32.2 $\pm$ 2.1 & 18.5 $\pm$ 3.4 & 0.25 $\pm$ 0.10 & 950 $\pm$ 380    \\ 
39.0 & 60.8  & 6073  & 21.0 $\pm$ 0.6 &  7.3 $\pm$ 2.8 & 0.10 $\pm$ 0.04 & 600 $\pm$ 240   \\
60.8 & 87.9  & 12278 & 16.7 $\pm$ 0.6 &  3.0 $\pm$ 2.8 & 0.04 $\pm$ 0.05 & 490 $\pm$ 610   \\

\enddata

\tablenotetext{a}{1 pix = 0.333\arcsec.}
\tablenotetext{b}{Estimated number of GCs in the central region
 of the galaxy. See text for details.}
\tablenotetext{c}{This galaxy is located in the NGC 4889 ring-shaped region with
$[r_{min},r_{max}]=[71.9 \arcsec, 104.7 \arcsec]$, so $\sigma_{\rm GC}^2$
of NGC 4889 region has been also subtracted to $P_0$.}

\end{deluxetable}

\clearpage
\begin{deluxetable}{lcccc|c}
\tabletypesize{\scriptsize}
\tablecaption{$\frac{\sigma^2_{\rm GC}}{N_{\rm GC}}$ results for different GCLF parameters. See text for details. \label{calibr_3}}
\tablewidth{0pt}
\tablehead{
\colhead{Galaxy} & \colhead{(27.22, 1.40)} & \colhead{(27.62, 1.40)} & \colhead{(27.42, 1.35)} & \colhead{(27.42, 1.45)} & \colhead{Adopted} 
}

\startdata

NGC 4874          & 0.088 & 0.051 & 0.060 & 0.075 & 0.067 $\pm$ 0.016 \\
NGC 4889          & 0.088 & 0.051 & 0.060 & 0.075 & 0.067 $\pm$ 0.016 \\
NGC 4886          & 0.088 & 0.051 & 0.060 & 0.075 & 0.067 $\pm$ 0.016 \\
IC 4012           & 0.088 & 0.051 & 0.060 & 0.075 & 0.067 $\pm$ 0.016 \\
IC 4021           & 0.088 & 0.051 & 0.060 & 0.075 & 0.067 $\pm$ 0.016 \\
IC 4026           & 0.088 & 0.051 & 0.060 & 0.075 & 0.067 $\pm$ 0.016 \\
IC 4041           & 0.088 & 0.051 & 0.060 & 0.075 & 0.067 $\pm$ 0.016 \\
IC 4045           & 0.088 & 0.051 & 0.060 & 0.075 & 0.067 $\pm$ 0.016 \\
IC 4051           & 0.088 & 0.051 & 0.060 & 0.075 & 0.067 $\pm$ 0.016 \\
IC 3976           & 0.058 & 0.033 & 0.039 & 0.049 & 0.045 $\pm$ 0.011 \\
IC 3959           & 0.076 & 0.044 & 0.052 & 0.065 & 0.059 $\pm$ 0.014 \\
MCG +5 --31 --063 & 0.076 & 0.044 & 0.052 & 0.065 & 0.059 $\pm$ 0.014 \\
NGC 4839          & 0.059 & 0.034 & 0.040 & 0.050 & 0.046 $\pm$ 0.011 \\
NGC 4840          & 0.059 & 0.034 & 0.040 & 0.050 & 0.046 $\pm$ 0.011 \\
NGC 4816          & 0.078 & 0.045 & 0.053 & 0.066 & 0.060 $\pm$ 0.014 \\
NGC 4673          & 0.087 & 0.050 & 0.059 & 0.074 & 0.067 $\pm$ 0.016 \\ 
IC 3651           & 0.087 & 0.050 & 0.059 & 0.074 & 0.067 $\pm$ 0.016 \\

\enddata

\end{deluxetable}

\clearpage

\begin{deluxetable}{llcccc}
\tabletypesize{\scriptsize}
\tablecaption{Summary of results \label{sn}}
\tablewidth{0pt}
\tablehead{\colhead{Galaxy}&
\colhead{$M_V^{\rm TOT}$}&\colhead{$R_{\rm NGC 4874}$ 
($\arcmin$)}& \colhead{$N_{\rm GC}^{\rm tot}$}&\colhead{$S_N$}
}

\startdata

NGC 4874 & --23.26 $\pm$ 0.11 & 0      & 18200  $\pm$ 4100  &  9.0 $\pm$ 2.2  \\
NGC 4889 & --23.45 $\pm$ 0.11 & 8.05   &  9600  $\pm$ 2600  &  4.0 $\pm$ 1.2  \\
NGC 4886 & --21.08 $\pm$ 0.10 & 7.43   &  1220  $\pm$  480  &  1.8 $\pm$ 1.8  \\
IC 4012  & --20.00 $\pm$ 0.11 & 10.68  &   620  $\pm$  410  &  6.2 $\pm$ 4.1  \\
IC 4021  & --20.09 $\pm$ 0.10 & 10.88  &   160  $\pm$  160  &  1.5 $\pm$ 1.5  \\
IC 4026  & --20.35 $\pm$ 0.15 & 12.62  &   650  $\pm$  190  &  4.7 $\pm$ 1.5  \\
IC 4041  & --20.59 $\pm$ 0.20 & 16.38  &   470  $\pm$  160  &  2.7 $\pm$ 1.1   \\
IC 4045  & --21.00 $\pm$ 0.10 & 19.85  &  1100  $\pm$  510  &  4.4 $\pm$ 2.1  \\
IC 4051  & --21.74 $\pm$ 0.15 & 19.95  &  6300  $\pm$ 1300  & 12.7 $\pm$ 3.2  \\
IC 3976  & --20.24 $\pm$ 0.15 & 6.78   &  920   $\pm$  250  &  7.4 $\pm$ 2.2  \\
IC 3959  & --20.95 $\pm$ 0.16:\tablenotemark{a} & 12.67 & 880  $\pm$ 330  & 3.7 $\pm$ 1.5 \\
MGC +5 --31 --063 & --20.55 $\pm$ 0.14:\tablenotemark{a} & 8.93 & 2150 $\pm$ 650 & 13.0 $\pm$ 4.2 \\
NGC 4839 & --22.88 $\pm$ 0.10 & 43.12  & 10000   $\pm$ 2500  & 7.0  $\pm$ 1.9  \\
NGC 4840 & --21.24 $\pm$ 0.13 & 37.43  & 1290   $\pm$ 620   & 4.1  $\pm$ 2.0  \\
NGC 4816 & --22.10 $\pm$ 0.20 & 52.60  & 6100   $\pm$ 1400  & 8.8 $\pm$ 2.6  \\
NGC 4673 & --22.14 $\pm$ 0.18:\tablenotemark{a} & 217.07 & 750  $\pm$ 290 & 1.0 $\pm$ 0.4 \\
IC 3651  & --21.68 $\pm$ 0.14:\tablenotemark{a} & 290.32 & 2700 $\pm$ 780 & 5.7 $\pm$ 1.8  \\

\enddata
\tablenotetext{a}{Estimated magnitude. See text for details.}

\end{deluxetable}

\end{document}